\documentclass[useAMS,usenatbib]{mn2e}
\usepackage{graphicx}
\usepackage{epstopdf}
\usepackage{amsmath}
\usepackage{amssymb}
\usepackage{enumerate}
\usepackage{subfigure}
\usepackage{bbding}
\usepackage[T1]{fontenc}

\usepackage{bm}		
\usepackage{booktabs,caption}
\usepackage[flushleft]{threeparttable}

\usepackage{hyperref}
\hypersetup{colorlinks,allcolors=blue}
\title[Elliptical galaxies with cosmological inflow]{Active galactic nuclei feedback in an elliptical galaxy (III): the impacts and fate of cosmological inflow}
\author[B. Zhu et al.]
{Bocheng Zhu$^{1,2}$, Feng Yuan$^{1,2}$\thanks{E-mail: fyuan@shao.ac.cn}, Suoqing Ji$^1$,  Yingjie Peng$^{4,3}$, Luis C. Ho$^{3,4}$, 
\newauthor 
Jeremiah P. Ostriker$^{5,6}$, and Luca Ciotti$^{7}$ \\
$^{1}$ Key Laboratory for Research in Galaxies and Cosmology, Shanghai Astronomical Observatory, Chinese Academy of Sciences, \\ 80 Nandan Road, Shanghai 200030, People's Republic of China \\
$^{2}$ School of Astronomy and Space Sciences, University of Chinese Academy of Sciences, No. 19A Yuquan Road, Beijing 100049,\\ People's Republic of China \\
$^{3}$ Kavli Institute for Astronomy and Astrophysics, Peking University, 5 Yiheyuan Road, Beijing 100871, People's Republic of China\\
$^{4}$ Department of Astronomy, School of Physics, Peking University, 5 Yiheyuan Road, Beijing 100871, People's Republic of China\\
$^{5}$ Department of Astronomy, Columbia University, 550 W, 120th Street, New York, NY10027, USA\\
$^{6}$ Department of Astrophysical Sciences, Princeton University, Princeton, NJ 08544, USA\\
$^{7}$ Department of Physics and Astronomy "Augusto Righi", University of Bologna, via Gobetti 93/2, I-40129 Bologna, Italy}
\begin{document}

\pagerange{\pageref{firstpage}--\pageref{lastpage}} \pubyear{2002}

\maketitle

\label{firstpage}
\begin{abstract}

The cosmological inflow of a galaxy is speculated to be able to enter the galaxy and enhance the star formation rate (SFR) and black hole accretion rate (BHAR). In this paper, by performing high-resolution hydrodynamic simulations in the framework of {\it MACER}, we investigate the fate of the inflow and its impacts on the evolution of a massive elliptical galaxy. The inflow properties are adopted from the cosmological simulation IllustrisTNG. We find that the inflow gas hardly enters but is blocked beyond $\sim20$ kpc from the central galaxy and becomes part of the circumgalactic medium (CGM). The gas pressure gradient, mainly contributed by the thermalized stellar wind and subdominant contributed by the energy input from the AGN, balances gravity and prevents the inflow from entering the galaxy. The SFR and BHAR are almost not affected by the normal inflow. However, if the rate of cosmological inflow were increased by a factor of 3, a small fraction of the inflow would enter the galaxy and contribute about 10\% of the gas in the galaxy. In this case, the gas density in the galaxy would increase by a factor of $\ga$ 20. This increase is not because of the additional gas supply by the inflow but due to the increase of gas density and pressure in the CGM caused by the inflow. Consequently, the SFR and BHAR would increase by a factor of $\sim5$ and $\sim1000$, respectively. Finally, AGN feedback can perturb the motion of the inflow and heat the CGM through its intermittent outbursts.

\end{abstract}


 \section{Introduction}

One of the puzzles in galaxy evolution is the interplay between galaxies and cosmological inflow. Many details about how the cosmological inflow affects galaxy evolution are still lacking. In classical theory, the gas inflow comes from the spherical collapse due to the gravity of dark matter halo \citep{Rees77, white78}. The inflowing gas following the collapse of the dark matter will experience a virial shock and become a spherical hot gas inflow. However, numerical simulations found that spherical virialized hot gas inflow and cold streams/clumps may co-exist \citep{keres05,vdv11,nelson13}. The cold gas inflow into the halo is usually not spherically symmetric but in the form of filaments or clumps and does not need to experience the virial shock. The cold gas inflow remains cold when penetrating the hot gaseous halo \citep{keres05, keres09, nelson13}. The mass ratio of the cold gas inflow to the total inflow increases with decreasing halo masses and increasing redshifts, so the cold inflow is the dominant gas accretion mode in high redshift and low-mass galaxies.

Over the past decades, cosmological simulations have shown the presence of inflow from large-scale structures and its interaction with feedback processes from galaxies, which must play a vital role in the galaxy evolution \citep{nelson15, correa18a, correa18b}. The gas inflow is rapid if the halo mass is small and the virial radius is smaller than the cooling radius. In a massive halo, if the virial radius is larger than the cooling radius, the gas inflow will be shock-heated to the virial temperature, and the gas supply will become slow.

However, the fate and impacts of cosmological inflow on the galaxy evolution, especially in galaxies larger than $10^{13}~\mathrm{M_{\odot}}$ at $z=0$, are still poorly understood. Cosmological simulations show that the gas inflow rate can exceed $100~\mathrm{M_{\odot}\ yr^{-1}}$ in massive galaxies at low redshift \citep{keres05, dekel09, vdv11, nelson13}. Although the inflowing gas is dominated by virialized gas and vulnerable to energetic feedback, it is still a potential fuel for star formation. On the other hand, most massive galaxies in the present day are quenched. Cosmological simulations have found that the cold stream inflow may be a significant gas resource for star formation in the formation stage of the massive halo at high redshifts \citep{dekel09}. \citet{nelson15} used the Illustris simulation suite \citep{vogelsberger14} to study the interaction between feedback and cosmological inflow and found that the feedback processes can suppress the inflow rate.  \citet{correa18b} also used the EAGLE simulation suite \citep{crain15} to study the interaction between inflow and feedback and { found} that the feedback can strongly suppress the inflow rate at $\sim10^{12}~\mathrm{M_{\odot}}$ halo. However, these works do not focus on the impacts of cosmological inflow on the evolution of massive galaxies at low redshifts. 

Many idealized galaxy simulation works have been done to analyze the reason for quiescence, usually by invoking active galactic nuclei (AGN) feedback \citep[e.g.,][]{gaspari12, li15, prasad15, 2017ApJ...835...15C,yuan18, wang19}. However, in these works, cosmological inflow is not considered. Some works also have discussed the interplay between cosmological inflow and massive galaxies. \cite{dekel06} and \cite{birnboim07} have shown that the virial shocks in the massive galaxies at low redshifts can heat the inflowing gas and make them vulnerable to the energetic feedback unless galaxy halos are dominated by cosmic ray energy where virial shocks are suppressed \citep{ji2020properties,ji2021virial}. In analytical work, \cite{voit20} have proposed that the virialized gas inflow in massive galaxies may reside in the CGM and provide the pressure to control the thermal state of the gas in galaxy regions, then further control the valve of AGN feedback. In this work, AGN feedback is not included. Some important questions, such as the impacts of cosmological inflow for the quenched galaxies, still need to be answered. 

In the present work, by performing high-resolution numerical simulations, we focus on the fate and impact of cosmological inflow in galaxy evolution, with AGN and stellar feedback included. We will try to answer the following questions: 1) can cosmological inflow enter the galaxy? 2) if it cannot, what is the physical mechanism for stopping it? 3) why can massive galaxies remain quenched even in the presence of cosmological inflow, and how does it affect star formation in the galaxy? 4) whether and how can the AGN feedback affect the cosmological inflow? 

The simulations are performed in the framework of {\it MACER}, with developments on stellar yields and the AGN physics at the super-Eddington region in the present paper. Briefly speaking, {\it MACER} is an idealized elliptical galaxy simulation framework developed based on early works \citep{ciotti01,ciotti09,ciotti10, Ostriker10,novak11,gan14}. In the most updated version, \citet{yuan18} have incorporated the state-of-the-art AGN physics, including radiation and wind as a function of accretion rates in both the hot and cold accretion (feedback) modes. The effects of AGN feedback on the black hole growth, AGN light curve, and star formation are discussed and compared to observations. \citet{yoon18}  extended \citet{yuan18} to the case of elliptical galaxies with a large angular momentum. The respective roles of AGB heating, supernovae feedback, and AGN feedback in the evolution of galaxies were investigated in \citet{Li18}.   The role played by the hot mode in the feedback was discussed in \citet{yoon19}, and it was found that the total mass of newly formed stars would be two orders of magnitude smaller if we only adopted the cold mode no matter what value the accretion rate is. 

The structure of the paper is as follows. In section \ref{model}, we review the key features of the {\it MACER}, especially the AGN physics adopted, and the new developments to {\it MACER} we have made in this work. We describe the main inflow properties and how they are extracted from the IllustrisTNG cosmological simulations in Section \ref{inflowdata}. Section \ref{setup} introduces the setup of our fiducial and reference models. Our results are presented in detail in Section \ref{results}, trying to answer questions like whether the cosmological inflow can enter the galaxy and why it is blocked, the effect of inflow on SFR and BHAR in the galaxy, and whether and how AGN feedback can affect the inflow and CGM. We finally conclude in Section \ref{summary}.

\section{Physics included in {\it MACER} }\label{model}

\subsection{Key features of {\it MACER}}

The {\it MACER} framework we will use in this paper is based on \cite{yuan18}. We first briefly introduce its several key features below.  

One is that the inner boundary of the {\it MACER} simulation domain is small enough to resolve the outer boundary of the accretion flow, i.e., the Bondi radius. In the case of a massive elliptical galaxy, \cite{yao21} have calculated the value of the Bondi radius and found that it is typically ten times larger than the inner boundary. Once the mass rate at the inner boundary is calculated, we can combine this value with the theory of black hole accretion to obtain the accretion at the black hole horizon. This accretion rate determines the power of the AGN and is the most crucial parameter to determine AGN feedback  \citep[see also,][]{angeles21}. In contrast,  cosmological simulations have a much poorer resolution. Thus it is difficult to determine the exact value of the black hole accretion rate. The spatial resolution of {\it MACER} achieved at the inner region of the simulation domain, where the interaction between AGN outputs and interstellar medium is the strongest, is as high as 0.2 pc in \citet{yuan18}.   

The second advantage of the {\it MACER} is that the AGN physics adopted in the code has considered the latest developments in black hole accretion theory. These mainly include the wind in both the cold and hot accretion (feedback) modes and the radiation in the hot mode. For example, the properties of the wind in the hot mode, such as the velocity and mass flux, are taken from \citet{yuan15}, as will be detailed in section \ref{agnphysics}. Given that the hot mode plays an essential role in AGN feedback \citep{yoon19} and the wind is more important than radiation in suppressing star formation, such a correct treatment is essential.

The third advantage is the exact calculation of the interaction between AGN outputs (wind and radiation) and the gas in the galaxy. We do not adopt the parameterized approach as usually adopted in most feedback works, e.g., assuming some percentage of the AGN power is deposited into an assumed region surrounding the AGN. For both radiation and wind, we directly inject them at the inner boundary of the simulation domain and calculate self-consistently their energy and momentum interaction with the ISM.   

Compared with \citet{yuan18}, we have added some new physics into the model in the present work, as we will introduce below. In the following part of this section, for the convenience of readers, we introduce the main physics adopted in \citet{yuan18}, plus the new physics added in this work. 

\subsection{AGN physics}
\label{agnphysics}
Black hole accretion is divided into cold and hot modes, bounded by $2\%L_{\rm Edd}$ in terms of the bolometric luminosity, or $\dot{M}_{\rm BH}\sim2\%\dot{M}_{\rm Edd}$ in terms of the mass accretion rate at the black hole horizon (here $\dot{M}_{Edd}\equiv 10 L_{\rm Edd}/c^2$) \citep{yuan14}. The critical accretion rate is based on the observations of the transition of the black hole X-ray binaries between hard and soft states \citep{mc06}. We believe that the value of the critical luminosity should be independent of the black hole mass. In our model, we first calculate the mass flux at the inner boundary of the simulation $\dot{M}(r_{\rm in})$. We then judge whether the accretion is in the cold or hot modes according to the comparison between $\dot{M}(r_{\rm in})$ and $2\%\dot{M}_{\rm Edd}$. The cold mode accretion is further divided into the standard thin disk \citep{shakura73} and super-Eddington accretion, bounded by $\dot{M}_{\rm Edd}$. Note that there is no so-called ``Eddington limit'', and the accretion rate can be much higher than the Eddington accretion rate, as indicated by both theoretical studies \citep[e.g.,][]{1988ApJ...332..646A,2005ApJ...628..368O,2014MNRAS.439..503S,2014ApJ...796..106J} and observational ones \citep[e.g.,][]{2013ApJ...764...45K}. In the hot accretion mode, we have three kinds of output: wind, jet, and radiation.  In the present work, we neglect jet, which is a caveat. We will investigate the role of jet in our future work. However, in the cold mode, we assume only to have wind and radiation\footnote{Observations find that about 10\% of the quasars are radio loud, which suggests that sometimes jets still exist even though the accretion is in the cold mode. The physics is still not understood, so we temporarily assume in the paper that there is no jet in the cold mode.}. The physics of radiation and wind in the two modes are completely different. 

 When the accretion rate is higher than $2\%\dot{M}_{\rm Edd}$, the AGN is in the cold mode. The inflow through the inner boundary first freely falls until a disk is formed
at the circularization radius. The black hole accretion rate $\dot{M}_{\rm BH}$ is calculated from $\dot{M}(r_{\rm in})$ by solving the following set of differential equations, taking into account the mass evolution of the small disk and the mass lost in the wind,

\begin{equation}
\frac{d\dot{M}_{\rm eff}}{dt} = \frac{\dot{M}(r_{\rm in})-\dot{M}_{\rm eff}}{\tau_{\rm ff}},
\end{equation}
\begin{equation}
M_{\rm dg} = \int \dot{M}_{\rm eff} dt,
\end{equation}
\begin{equation}
\dot{M}_{\rm d, inflow} = \frac{M_{\rm dg}}{\tau_{\rm vis}},
\end{equation}
\begin{equation}
\dot{M}_{\rm BH} = \dot{M}_{\rm d, inflow}-\dot{M}_{\rm wind}.
\end{equation}
Here $\dot{M}_{\rm eff}$ is the effective accretion rate that the gas falls into the small disk, $\tau_{\rm ff}\equiv r_{\rm in}/(2GM_{\rm BH}/r_{\rm in})^{1/2}$ is the free-fall time scale from the inner boundary to the small disk, $M_{\rm dg}$ is the total mass of small disk, $\tau_{\rm vis}\equiv 1.2\times10^6~(M_{\rm BH}/10^9~{\rm M_{\odot}})~ \mathrm{yr}$ is the instantaneous viscous timescale, $\dot{M}_{\rm d, inflow}$ is the accretion rate from the small disk to the accretion disk, and $\dot{M}_{\rm wind}$ is the mass loss rate via the disk wind. The calculation of the mass flux of the wind will be given in the following paragraph.

The bolometric luminosity is calculated by $L_{\rm bol}=\epsilon_{\rm thin} \dot{M}_{\rm BH}c^2$, with $\epsilon_{\rm thin}$ is the radiative efficiency of the standard thin disk and is set to be 0.1. Although there have been numerous theoretical studies on the wind launched from a thin disk\citep[e.g.,][]{2022MNRAS.513.5818W}, since the observational data is very abundant, we directly use the statistical results of the mass flux and velocity of wind as a function of $L_{\rm bol}$ \citep{gofford15}. Following  \citet{yuan18}, the mass flux and velocity of the wind from cold mode can be described as
\begin{equation}
   \dot{M}_{\mathrm{wind,cold}} = 0.28\left(\frac{L_{\mathrm{bol}}}{10^{45} \mathrm{erg\ s^{-1}}}\right)^{0.85}~\mathrm{M_{\odot}~yr^{-1}},
\end{equation}
\begin{equation}
   v_{\mathrm{wind,cold}} = \min\left(2.5\times 10^4 \left(\frac{L_{\mathrm{bol}}}{10^{45} \mathrm{erg\ s^{-1}}}\right)^{0.4}, 10^5\right)\ \mathrm{km\ s^{-1}}.
\label{vcoldwind}
\end{equation}
In equation \ref{vcoldwind}, we set $10^5~\mathrm{km/s}$ as an upper limit of the wind since observations indicate that the velocity of the wind will be saturated at this value. We set the angle distribution of the wind mass flux as $\dot{M}(\theta)\propto \cos^2\theta$.

When $\dot{M}_{\rm BH}\ga \dot{M}_{\rm Edd}$, the accretion will be in the super-Eddington mode. This accretion regime is neglected in  \cite{yuan18} but is considered in the present work. 
Specifically, the mass flux and velocity of wind as a function of $\dot{M}_{\rm BH}$ and radius will be taken from \cite{2023MNRAS.523..208Y}. In this work, they have performed three-dimensional general relativistic radiation
magnetohydrodynamical (RMHD) numerical simulations of a super-Eddington accretion flow around a black hole and analyze the data using the ``virtual particle trajectory'' approach, which can loyally reflect the motion of fluid elements and discriminate turbulence and real wind, to obtain the wind properties. The radiative efficiency as a function of $\dot{M}_{\rm BH}$ is taken from fitting the three-dimensional RMHD simulation data of \citet{jiang19}. The mass flux and velocity of wind and the radiative efficiency of the super-Eddington accretion flow are described by:
\begin{equation}
    \dot{M}_{\mathrm{wind,super}} = \left(\frac{r_{\rm d}}{45r_{\rm s}}\right)^{0.83}\dot{M}_{\mathrm{BH}},
\end{equation}
\begin{equation}
    v_{\mathrm{wind,super}} = 0.15 c
\end{equation}
\begin{equation}
    \epsilon_{\mathrm{super}} = 0.21\left(\frac{100\dot{M}_{\mathrm{BH}}}{\dot{M}_{\mathrm{Edd}}}\right)^{-0.17}.
\end{equation}
Here $r_{\rm d}$ is the outer boundary of the super-Eddington accretion flow. The angle distribution of the mass flux is set to $0^{\circ}-30^{\circ}$ and $150^{\circ}-180^{\circ}$.

When the accretion rate is lower than $2\%\dot{M}_{\mathrm{Edd}}$, the accretion is in hot mode. The accretion flow consists of a truncated cold disk outside $r_{\rm tr}$ and a hot accretion flow within this radius \citep{yuan14}. The truncation radius is described by
\begin{equation}
    r_{\mathrm{tr}} \approx 3 R_{\mathrm{s}}\left[\frac{2\times10^{-2}\dot{M}_{\mathrm{Edd}}}{\dot{M}(r_{\rm in})}\right]^2\label{con:rtr},
\end{equation}
where $R_{\mathrm{s}}$ is the Schwarzschild radius.
MHD numerical simulations have shown the existence of strong wind in hot accretion flows \citep{2012ApJ...761..130Y,2012MNRAS.426.3241N,yuan15}. Using three-dimensional general relativity MHD numerical simulation data and ``virtual test particle trajectory'' approach, the properties of wind as a function of accretion rate and black hole spin have been obtained \citep{yuan15,2021ApJ...914..131Y}. On the observational side, we are accumulating more and more observational evidence for wind from hot accretion flows \cite[e.g.,][]{wang13,cheung,2019ApJ...871..257P,2019MNRAS.483.5614M,2021NatAs...5..928S}. However, these observations still can not provide a good constraint on the properties of the wind. So in this paper, following \cite{yuan18}, we adopt the results obtained in \cite{yuan15}. The black hole accretion rate, mass flux and velocity of the wind from hot mode are described as
\begin{equation}
   \dot{M}_{\mathrm{BH}} = \dot{M}(r_{\rm in})\left(\frac{3r_{\mathrm{s}}}{r_{\mathrm{tr}}}\right)^{0.5}.
\end{equation}
\begin{equation}
   \dot{M}_{\mathrm{wind,hot}} = \dot{M}(r_{\rm in})\left[1-\left(\frac{3r_s}{r_{\mathrm{tr}}}\right)^{0.5}\right]\label{con:hotwind},
\end{equation}
\begin{equation}
   v_{\mathrm{wind,hot}} = (0.2-0.4)v_{\mathrm{K}}(r_{\mathrm{tr}}),
\end{equation}
where $v_{\mathrm{K}}(r_{\mathrm{tr}})$ is the Keplerian velocity at truncated radius. Based on the analysis of \cite{yuan15}, the angle distribution of the wind is set to $30^{\circ}-70^{\circ}$ and $110^{\circ}-150^{\circ}$ in \cite{yuan18} and the present work. 

For the radiative efficiency of the hot accretion flow, our model has adopted the findings of \cite{xie12}. It is important to note that, unlike the standard thin disk, the efficiency of hot accretion flow is lower and dependent on the accretion rate. The radiation efficiency $\varepsilon_{\mathrm{hot}}$ can be described as
\begin{equation}
   \varepsilon_{\mathrm{hot}} (\dot{M}_{\mathrm{BH}})=\varepsilon_0\left(\frac{\dot{M}_{\mathrm{BH}}}{0.1L_{\mathrm{Edd}}/c^2}\right)^a.\label{con:eps}
\end{equation}
The values of parameter $\varepsilon_0$ and $a$ can be found in \cite{xie12}.

\subsection{Interaction of AGN outputs with ISM}
\label{interaction}

After the AGN outputs are obtained, we implement them into our simulations. For AGN wind, we treat it as a source term. We add the energy, momentum and mass of the wind into the innermost two grids of the simulation region. Then the energy, momentum and mass of the wind will automatically transport to a large scale in the simulations.

For AGN radiation, we consider the heating and radiation pressure. The radiation pressure is due to electron scattering and the absorption of photons by atomic lines. The radiative heating and cooling were calculated in \cite{sazonov05}. The radiative heating and cooling terms include bremsstrahlung, Compton heating and cooling,  photoionization heating, line and recombination continuum cooling. Compton temperature is required to calculate Compton heating/cooling. Its value is determined by the spectral energy distribution of the AGN. In the case of cold accretion mode, its value has been studied in \cite{sazonov05}, which is $T_{\rm C, cold}=2\times 10^7K$. In the case of hot accretion mode, the spectrum is completely different and much harder than the cold mode \citep{1999ApJ...516..672H}. The value of Compton temperature in this regime was calculated in \citet{xie17}, which is about ten times higher than $T_{\rm C, cold}$ and is adopted in {\it MACER}.

\subsection{Galaxy model}

Following previous work \citep{ciotti09, yuan18}, the dark matter halo and stellar distribution are set to a static, spherically symmetric distribution. The initial total stellar mass $M_{\star}$ is set to be $3\times 10^{11}\mathrm{M_{\odot}}$. The stellar distribution is adopted to the Jaffe \citep{jaffe83} profile:
\begin{equation}
    \rho_{\star} = \frac{M_{\star} r_{\rm J}}{4\pi r^2 (r_{\rm J}+r)^2}
\end{equation}
where $r_{\rm J}$ is the Jaffe radius. $r_{\rm J}$ is set to $6.9$~kpc in our simulations. The corresponding effective radius $r_{\rm e}$ is 9.04 kpc. The one-dimensional stellar velocity dispersion $\sigma_0$ is set to be $260\ \mathrm{km~s^{-1}}$. The effective radius $r_{\rm e}$ and the stellar velocity dispersion $\sigma_0$ are chosen to let the galaxy obey the fundamental plane \citep{djorgovski87} and Faber-Jackson relation \citep{faber76}. The black hole mass $M_{\mathrm{BH}}$ is set to be $1.8\times 10^9 \mathrm{M_{\odot}}$, which obeys the $M_{\mathrm{BH}}-\sigma$ relation \citep{kormendy13}. The total density profile is assumed to be an isothermal sphere. The dark matter mass and virial radius are $2\times10^{13}~\mathrm{M_{\odot}}$ and 513~kpc at $z=0$.

In this paper, we only consider galaxies with low angular momentum. Since the rotation of the galaxies is slow, we can calculate the stellar velocity dispersion $\sigma_{\star}$ by Jeans equation. The stellar velocity dispersion is an important parameter for calculating the thermalization of stellar mass loss, which is a significant part of stellar feedback in elliptical galaxies.

\subsection{Star formation and stellar feedback}

The model of star formation, stellar mass loss, and stellar feedback are based on a previous work \citep{ciotti12}, but we use the tabular mass loss $\Delta M(M_{\star}, Z_{\star})$ \citep{karakas10, doherty14a, doherty14b, nomoto13} in this work. The stellar feedback includes thermalized stellar wind, SN Ia, and SN II. Since the star formation rate is relatively low in elliptical galaxies, the stellar feedback is dominated by the thermalization of stellar wind and SN Ia feedback. For detailed descriptions of stellar mass loss and stellar feedback, please refer to \cite{ciotti12}. A brief introduction will be given below. Compared to \cite{ciotti12}, the stellar metallicity is considered, affecting the star's mass loss rate. We set a static stellar metallicity profile. We also consider the stellar yields. Like the mass loss, the tabular metal release $\Delta Z(M_{\star}, Z_{\star})$ is used. This part of the model will be introduced in section \ref{yields}.

\subsubsection{Star formation}

The star formation model is the same as the previous work, except that we consider the density and temperature thresholds for star formation, as widely adopted in numerical simulation works due to the insufficient resolutions. In this work, we require that gas can only be converted into stars if its density is higher than $1~\mathrm{cm^{-3}}$ and the temperature is lower than $4\times 10^4~\mathrm{K}$.

\subsubsection{Stellar mass loss and stellar feedback}

The same as previous work, the stellar mass-loss rate following the stellar evolution theory \citep{maraston05} can be described by 
\begin{equation}
    \dot{M}_{\star} = \mathrm{IMF}(M_{\mathrm{TO}})|\dot{M}_{\mathrm{TO}}|\Delta M \label{con:massloss}
\end{equation}
where $\mathrm{IMF}(M_{\star})$ is the initial mass function (IMF) and is set to Salpeter IMF, $M_{\mathrm{TO}}$ is the turn-off mass of the star, $\Delta M$ is the total mass loss of a star of $M_{\mathrm{TO}}$. Unlike previous work, we use the turn-off mass given in \cite{cristallo15}, which is a function of the age of the stellar population and the stellar metallicity. 

Following \citet{spolaor11}, we set the radial profile of star metallicity as:
\begin{equation}
    (Z/Z_{\odot})(r) = \exp[-0.23\log(r/r_{\rm e})]+0.3.
\end{equation}
The stellar mass loss rate can be calculated using equation \eqref{con:massloss}. 

The SN Ia is also the same as previous work. The SN Ia feedback energy is added to the simulation uniformly in the form of thermal energy. However, Recent works \citep{li20a, li20b} show that the inhomogeneity of the ISM due to the SN feedback will enhance the net heating rate. To take this effect into account, we increase the energy of Stellar Feedback by a factor of 1.4 (Miao Li, private communication).

Following previous work, the mass return due to the massive star and energy output from SN II is calculated by
\begin{equation}
    \dot{\rho}_{II} = \mathrm{SFR}\cdot\frac{\int_{8 \mathrm{M_{\odot}}}^{M_{\max}}\mathrm{IMF}(m)\Delta M(m,Z_{\star}) dm}{\int_{M_{\min}}^{M_{\max}}\mathrm{IMF}(m)m dm}\label{con:ii}
\end{equation}
\begin{equation}
    \dot{E}_{II} = \mathrm{SFR}\cdot E_{\mathrm{II}}\int_{8 \mathrm{M_{\odot}}}^{M_{\max}}\mathrm{IMF}(m) dm
\end{equation}
where $M_{\min}$ and $M_{\max}$ is the lower and upper limit of IMF, $E_{\mathrm{II}}$ is the energy released by single SN II, and $\Delta M(m, Z_{\star})$ is the total mass of the star during its evolution, which is a function of stellar mass and stellar metallicity. Here we assume that $M_{\max}$ is $40\ \mathrm{M_{\odot}}$, $M_{\min}$ is $0.5~\mathrm{M_{\odot}}$ and $E_{\mathrm{II}}$ is $10^{51}~\mathrm{erg}$.

\subsubsection{Stellar yields}\label{yields}

Since the metallicity will affect the gas cooling rate, we include the stellar yield in this work. Similar to the mass return from the old star described in equation \eqref{con:massloss}, we can calculate the metal release from the mass loss of the old star by 
\begin{equation}
    \dot{Z}_{\star} = \mathrm{IMF}(M_{\mathrm{TO}})|\dot{M}_{\mathrm{TO}}|\Delta Z
\end{equation}
where $\Delta Z$ is the total metal release of a star with mass $M_{\mathrm{TO}}$. 

By assuming IRA, the metal release from the mass return of a newly formed star is similar to equation \eqref{con:ii}, which is expressed as
\begin{equation}
    \dot{Z}_{\rm II} = \mathrm{SFR}\cdot\frac{\int_{8 \mathrm{M_{\odot}}}^{M_{\max}}\mathrm{IMF}(m)\Delta Z(m,Z_{\star}) dm}{\int_{M_{\min}}^{M_{\max}}\mathrm{IMF}(m)m dm},
\end{equation}
where $\Delta Z(m, Z_{\star})$ is the metal release as a function of the stellar mass and metallicity.

\begin{figure*}
    \includegraphics[width=0.9\textwidth]{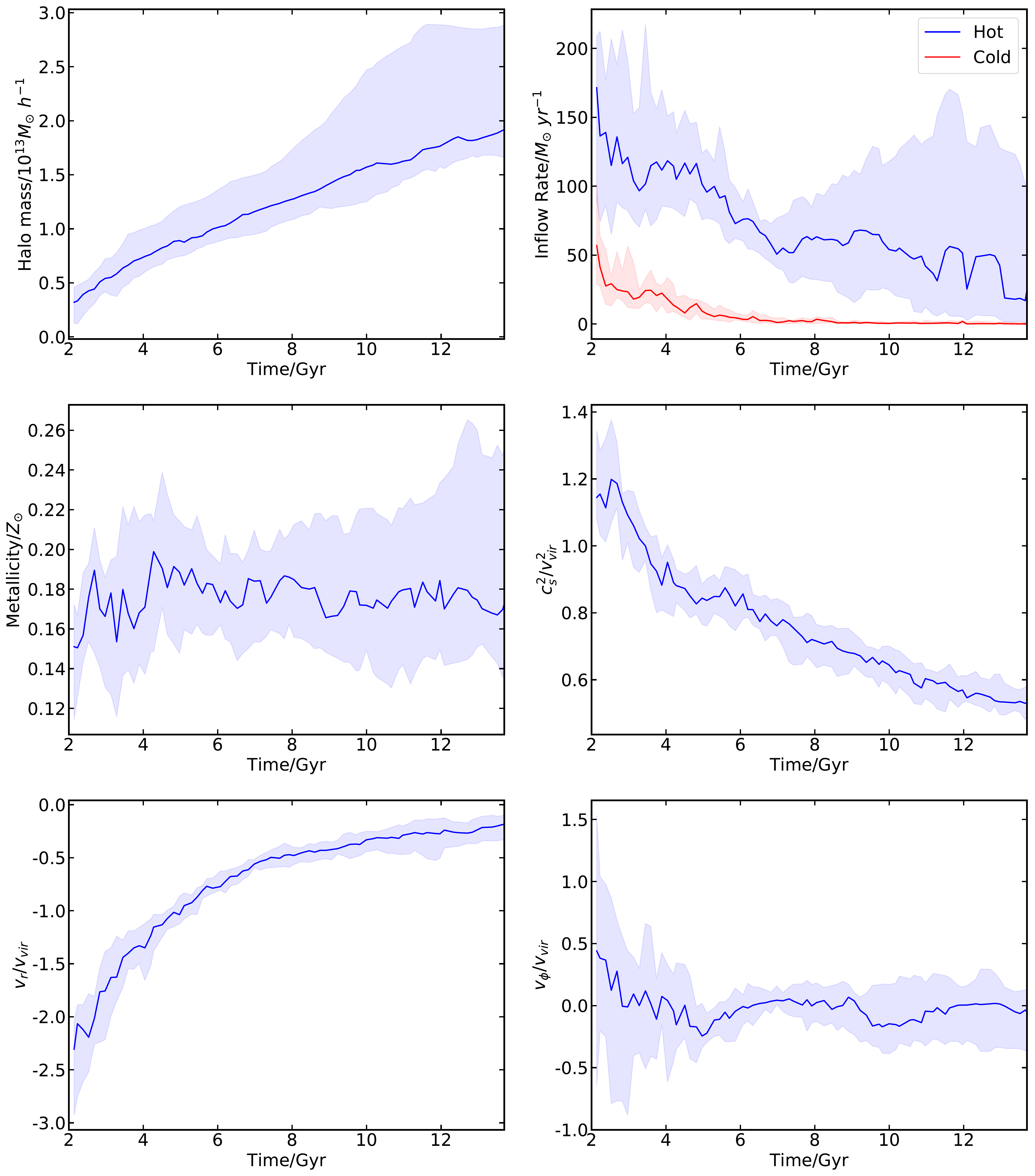}
    \caption{The time evolution of the median value of the halo mass, inflow rate, metallicity, square of sound speed, radial velocity, and rotational velocity of the hot gas inflow for the selected galaxies extracted from TNG100 data. The shaded areas represent the corresponding values in the 25th and 75th percentiles in selected galaxies.}
    \label{tng_inflow}
   \end{figure*}
   
\section{Cosmological inflow}
\label{inflowdata}

Several previous works have included the cosmological inflow based on the {\it MACER} framework \citep{gan19, gan20, ciotti22}. However, the fate and impacts of inflow on the galaxy evolution have not been fully investigated in those works. In the present paper, we use the IllustrisTNG simulation data to implement cosmological inflow into our simulations to investigate these questions. 

Many theoretical studies \citep{keres05, vdv11, nelson13} suggest that the cosmological inflow can be classified into two modes: the hot mode and the cold mode. For the hot mode, the gas inflow into the dark matter (DM) halo is close to the virial temperature, and the angle distribution is approximately spherically symmetric. In contrast, the temperature of the inflow gas of cold mode is relatively low and is in the form of filaments. There exists a critical mass of galaxies above which the cosmological flow is dominated by the hot mode and vice versa. This critical mass is around $10^{12}\mathrm{M_{\odot}}$ \citep{keres05, correa18a}. Since we focus on massive galaxies above this critical halo mass, we only consider the hot inflow. We will simulate the cold filament flow in future work.

\subsection{Inflow data from TNG}

We extract the inflow gas property from the public data of IllustrisTNG\footnote{https://www.tng-project.org/data/} \citep{nelson19} to implement it into our simulations. The IllustrisTNG project is a suite of state-of-the-art cosmological magneto-hydrodynamical simulations, including several sub-grid models such as star-formation and stellar feedback, BH accretion and feedback, etc. The IllustrisTNG project was run using a moving-mesh code AREPO\footnote{https://arepo-code.org/} \citep{springel10}. The TNG project includes three different simulations with different resolutions and simulation boxes: TNG50, TNG100, and TNG300. We extract the inflow gas data in dark matter halos of comparable mass to the ones we simulate from TNG100 \citep{springel18, pillepich18b, naiman18, nelson18, marinacci18}.

\subsection{Determining the gas accretion history}

To determine the gas accretion history of a DM halo in a cosmological simulation, one should identify a single dark matter halo and its merger tree. The public data of IllstrisTNG provide friends-of-friends (FoF) halo catalogues and use SUBFIND algorithm \citep{springel01} to identify the subhalos within FoF halos. For the subhalos presented in the simulation, IllstrisTNG provides two distinct merger trees in the public data: SUBLINK \citep{rg15} and LHALOTREE \citep{springel05}. In this work, we use the SUBLINK merger tree and trace back the gas accretion history of the main leaf progenitors.

Some previous works \citep{keres05, vdv11, correa18b} measure the gas accretion rate by counting the gas elements in current subhalos but not in its previous progenitors. It is a good scheme when we want to investigate the inflow gas properties and gas accretion history. However, such a method is unsuitable for us to extract the inflow gas information and add it to our grid-based code as a boundary condition. In this work, we use a scheme similar to the one described in \cite{fg11}. 
The gas accretion or outflow is determined within a fixed radius. Once this radius is selected, we count all the gas elements within a ring with a thickness of $\Delta_{\rm p}$ around the radius. Then the inflow rate is calculated by:
\begin{equation}
    \dot{M}_{\rm in} = \Sigma_{\rm p} M_{\rm p} v_{\rm p}(<0)/\Delta_{\rm p}.
\end{equation}
We repeat calculating the inflow rate by increasing $\Delta_{\rm p}$ until $\dot{M}_{\rm in}$ converges. The gas accretion rate calculated in this scheme is instantaneous.

\subsection{Inflow gas properties}

Since our work focuses on the evolution of 2 Gyr-old massive elliptical galaxies in which most of the stars have already formed, we add the following filters to select the halos from TNG100 simulation data:
\begin{enumerate}[1.]
\item The halo mass should be $\sim10^{13-14}~\mathrm{M_{\odot}}$ at $z=0$, which is similar to our simulated galaxy.

\item The mass ratio of halo at $z=1.6$ and $z=0$ should be less than 5. Since our simulated galaxy potential is static, we select the DM halos that do not grow too rapidly.

\item The ratio of stellar mass at $z=1.6$ and $z=0$ should be less than 2. Since we presume the simulated galaxies already formed at 2 Gyr after the big bang, we add this filter to ensure that the selected galaxies do not gain much more stars from that time on.

\item The central velocity dispersion should be $200-300~\mathrm{km~s^{-1}}$ at $z=0$.
\end{enumerate}

After adding the abovementioned filters, we selected 20 galaxies from TNG100 data. The inflowing gas properties are shown in Figure \ref{tng_inflow}. The dark matter halo masses in selected halos grow from $\sim5\times10^{12}~\mathrm{M_{\odot}}$ to $\sim2\times10^{13}~\mathrm{M_{\odot}}$. Since our simulations set a static dark matter halo, this setup may be slightly inconsistent with the accretion and growth of dark matter. The figure also shows that the inflow rate, metallicity, sound speed, radial velocity, and rotational velocity of the inflowing gas change with the cosmic age. However, the range of change is not very large. For simplicity, we assume that the properties of inflowing gas do not change with time. Since the cold inflow rate is much lower than the hot inflow, as shown in Figure \ref{tng_inflow}, we only consider the spherical hot inflow. In our fiducial model, the inflow rate, radial velocity, square of sound speed, metallicity, and rotational velocity are set to be 100~$\mathrm{M_{\odot}~yr^{-1}}$, 0.5~$v_{\mathrm{vir}}$, 0.8~$v_{\mathrm{vir}}^2$, $0.18~\mathrm{Z_{\odot}}$, and 0, respectively, where $v_{\rm vir}$ is the virial velocity of DM halo and $\mathrm{Z_{\odot}}$ is the solar metallicity.

\begin{table}
 \caption{The descriptions of different models}
 \label{tab:models}
 \begin{tabular*}{\columnwidth}{@{}l@{\hspace*{15pt}}l@{\hspace*{15pt}}l@{\hspace*{15pt}}l@{}}
 \hline
 Model & AGN feedback & SN feedback & cosmological inflow\\
 \hline
 Fiducial & \checkmark & \checkmark & \checkmark \\
 NoAGN & \XSolidBrush & \checkmark & \checkmark \\
 NoSN & \checkmark & \XSolidBrush & \checkmark \\
 NoFB & \XSolidBrush & \XSolidBrush & \checkmark \\
 NoInflow & \checkmark & \checkmark & \XSolidBrush \\
 3Inflow & \checkmark & \checkmark & $3\times$ inflow rate \\
 \hline
 \end{tabular*}
\end{table}

\begin{figure*}
    
    \includegraphics[width=\textwidth]{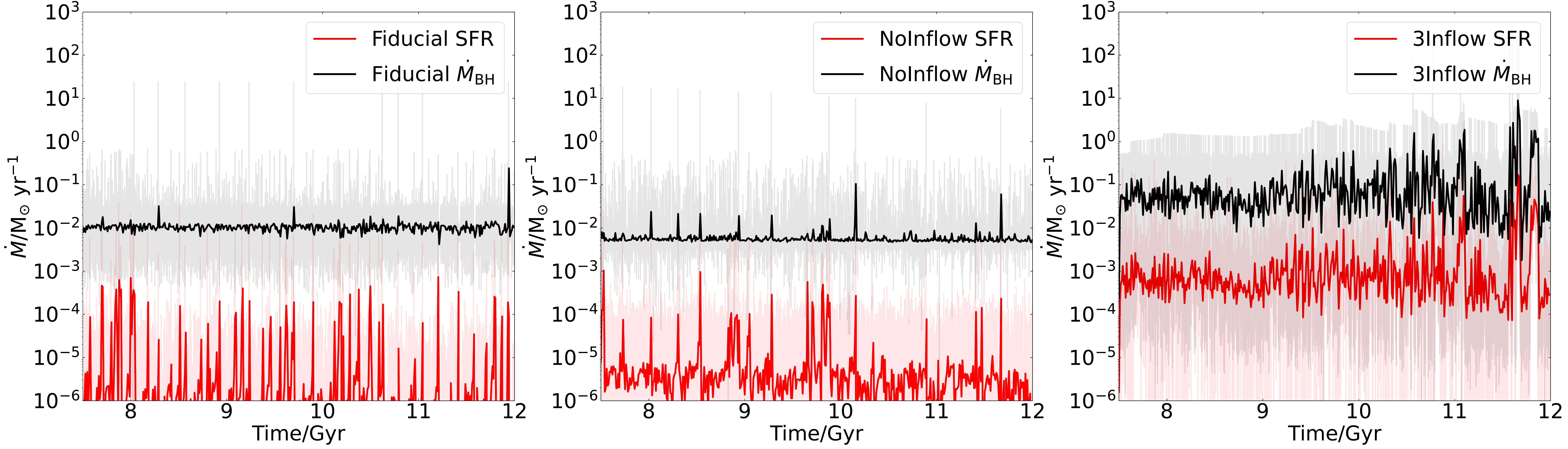}
    \caption{Time evolution of SFR (red lines) and BHAR (black lines) in the ``Fiducial'' ({\it left panel}), ``NoInflow'' ({\it middle panel}) and ``3Inflow" ({\it right panel}) models. The translucent lines represent data with a time interval of $10^4$ yr, while the opaque line represents data with a 100 Myr average.}
    \label{SFR&BHAR}
   \end{figure*}
   
\section{Setup of models}
\label{setup}

\subsection{Model setup}
The current {\it MACER} uses two-dimensional axisymmetric spherical coordinates ($r,\theta,\phi$). The inner and outer boundaries of the simulation domain are located at 2.5pc and 500 kpc, corresponding to less than the Bondi and virial radius, respectively. Our fiducial simulations have $240\times 60$ grids in the $r-\theta$ plane. The mesh is divided homogeneously in the $\theta$ direction, while we use a logarithmic mesh in the radial direction. A small range of $\theta$ around the axis is excluded to avoid the singularity. With such grids, the finest resolution is achieved in the innermost grid, $\sim$0.3 pc. Such a configuration ensures that the innermost region, where the AGN radiation and wind interact between AGN outputs and the ISM is the strongest, has the highest resolution. The radially inner and outer boundary conditions are set to the inflow-outflow boundary based on Roe Riemann solvers. At the pole, the boundary condition is set to symmetry. 

Unlike our previous works, we add gas as the initial condition to the simulation domain. The  distribution is set to be an $\beta$ model at large radii and a power law distribution at small radii:  
\begin{equation}
    \rho_{\mathrm{gas}}(r)=
    \begin{cases}
        \rho_0 (1+(r/r_{\rm c})^2)^{-3/2\beta} & r> r_{\rm c}\\
        \rho_0/2\ (r/r_{\rm c})^{\alpha} & r\le r_{\rm c}\\ 
    \end{cases}\label{con:dens}
\end{equation}
where $r_{\rm c}$ is set to be 6.9 kpc, $\beta$ to be $2/3$ based on observations \citep{anderson13}, and the power index $\alpha$ is set to be $-1$, which is consistent with some observations of the massive ellipticals \citep{wong14,russell18}. 

We set the initial baryon fraction in our simulated galaxies to be $50\%$ cosmic mean baryon fraction. Our setup for the gas profile has a more extended core than observations \citep{werner2012}. We find that the simulation will quickly turn to a quiescent state within $\sim$2 Gyr after the start of the simulation.

\subsection{Reference models}

To investigate the fate and impacts of cosmological flow in massive galaxies, we have run six models. The descriptions of these six models are listed in Table \ref{tab:models}. In the ``Fiducial'' model, all of the feedback processes and cosmological inflow described in section \ref{model} are adopted. The ``NoAGN'' model is the same as the ``Fiducial'' model, except that it does not include AGN feedback. Similarly, the ``NoSN'' model does not include SN feedback, while the ``NoFB'' model does not include both AGN and SN feedback. Finally, the ``NoInflow'' model does not include the cosmological inflow, while the ``3Inflow'' model includes three times of inflow rate adopted in the ``Fiducial'' model. Since the ``Fiducial'' model has a ``quenching'' phase at  $\sim$2 Gyr, we restart the ``Fiducial'' model at 6 Gyr to run ``NoAGN'', ``NoSN'', ``NoFB'', ``NoInflow'', and at 7.5 Gyr to run ``3Inflow'' models.

\section{Results}\label{results}

In section \ref{gr}, we present an overview of the evolution of the galaxy of the ``Fiducial'' model, including the evolution of the black hole accretion rate (BHAR) and star formation rate (SFR). In section \ref{inflow}, we focus on the trajectory of the cosmological inflow and its spatial distribution. We find that the inflow cannot enter the galaxy but stops at $\sim20$ kpc. To study the physical reasons why the inflow will stop, we investigate the ratio of the gas pressure gradient and gravity and the radial mixing in section \ref{why}. Section \ref{infb} further investigates how the gas pressure is established. We especially discuss whether the AGN and stellar feedbacks are essential in this process. Section \ref{effects} discusses the effects of cosmological inflow on the star formation rate and black hole accretion rate in the galaxy. Finally, in section \ref{AGN&inflow}, we discuss the impacts of AGN feedback on the cosmological inflow.

\subsection{Overview of the ``Fiducial'' model}
\label{gr}

\begin{figure*}
    \includegraphics[width=\textwidth]{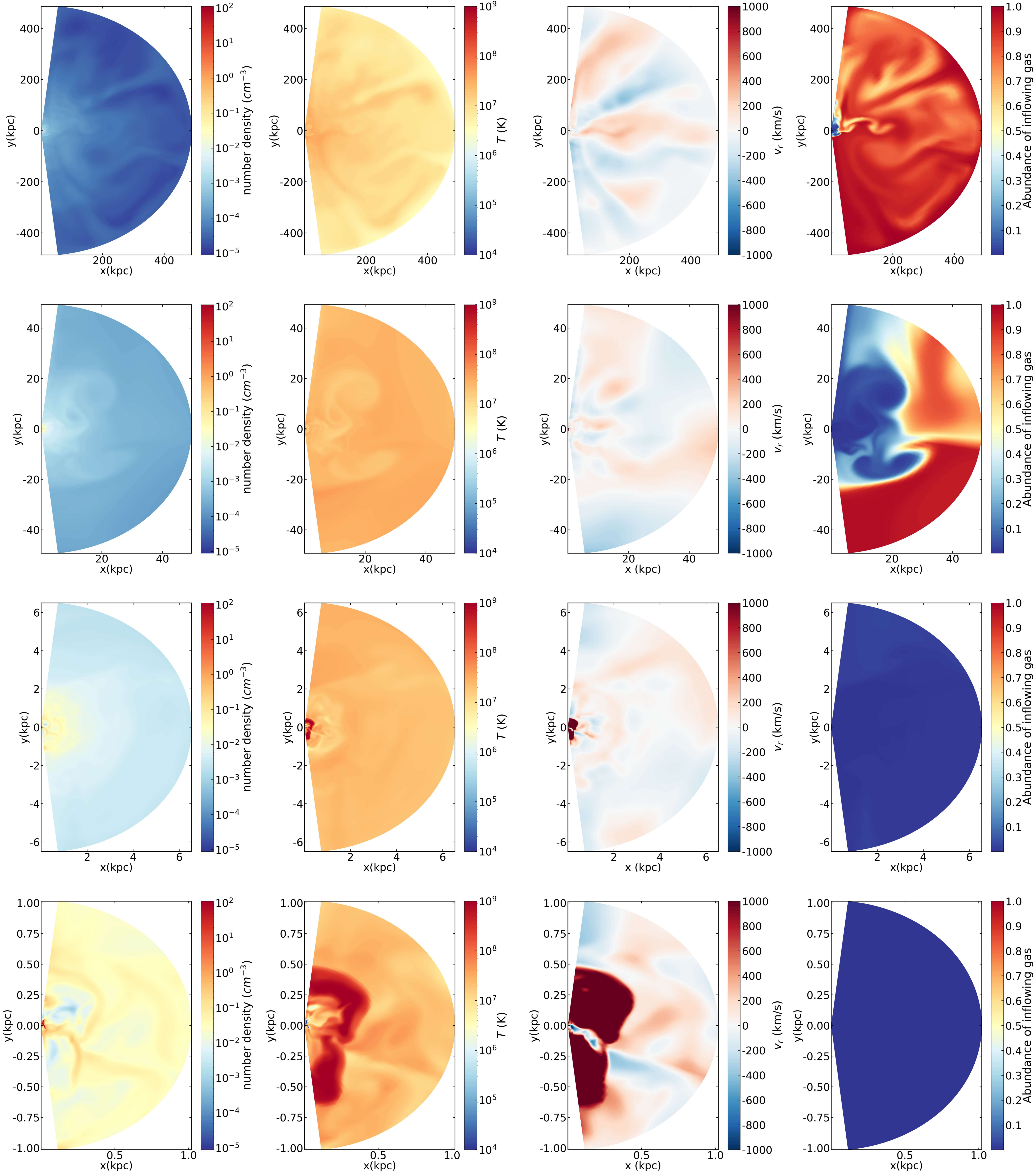}
    
    \caption{Snapshots of gas density, temperature, radial velocity, and inflow gas abundance at t = 7.875 Gyr with different scales for the ``Fiducial'' model. The inflow gas abundance indicates the proportion of the gas coming from the cosmological inflow. From top to bottom, each row represents the snapshot of various physical quantities within the virial radius, 5$r_{\rm e}$, $r_{\rm e}$, and 1 kpc, respectively.}
    \label{snap_quiet}
   \end{figure*}

The left panel of Figure \ref{SFR&BHAR} shows the time evolution of star formation rate (SFR) and black hole accretion rate (BHAR) in the ``Fiducial'' model from 7.5 to 12~Gyr. The middle and right panels show the results of the other two models; we will discuss them later. From the figure, we can see that the BHAR oscillates around $10^{-2}~\mathrm{M_{\odot}\ yr^{-1}}$($\sim10^{-3}~\dot{M}_{\mathrm{Edd}}$) with an oscillation amplitude fluctuating over three orders of magnitude. The star formation activity is also intermittent, and the value of SFR is typically much lower than $10^{-3}~\mathrm{M_{\odot}\ yr^{-1}}$. Such an SFR is far below the star-forming main sequence \citep[SFMS;][]{noeske07}. 

Figure \ref{snap_quiet} shows the snapshot of the gas number density, temperature, radial velocity, and inflow gas abundance of the ``Fiducial'' model at different scales at 7.875 Gyr when an AGN outburst occurs. We can see that the distributions of these quantities are quite inhomogeneous. The velocity distribution shows some complicated inflow-outflow fountain-like structures. From the third row of the figure, we can see that a strong outflow from the AGN leads to many hot bubbles and ripples. A forward shock is present at $r\sim2$ kpc, possibly the relic of a former outburst. Some large eddies appear at larger radii,  relics of AGN outbursts in the further past. From the fourth row of the figure, a strong AGN wind produces an evident shock. The front of a forward shock is evident at $r\sim0.5$ kpc. Inside the forward shock is a thick high-temperature layer of gas, which is shock-heated ISM. We also find the formation of some cold gas, likely due to the lifting of low-entropy gas from the center \citep[][]{li15,voit17} or the growth of local thermal instability \citep{mc11,sharma12}. 

\subsection{How deep can the cosmological inflow enter the galaxy? Simulation results of the ``Fiducial'' model}\label{inflow}

\begin{figure*}
    \includegraphics[width=0.9\textwidth]{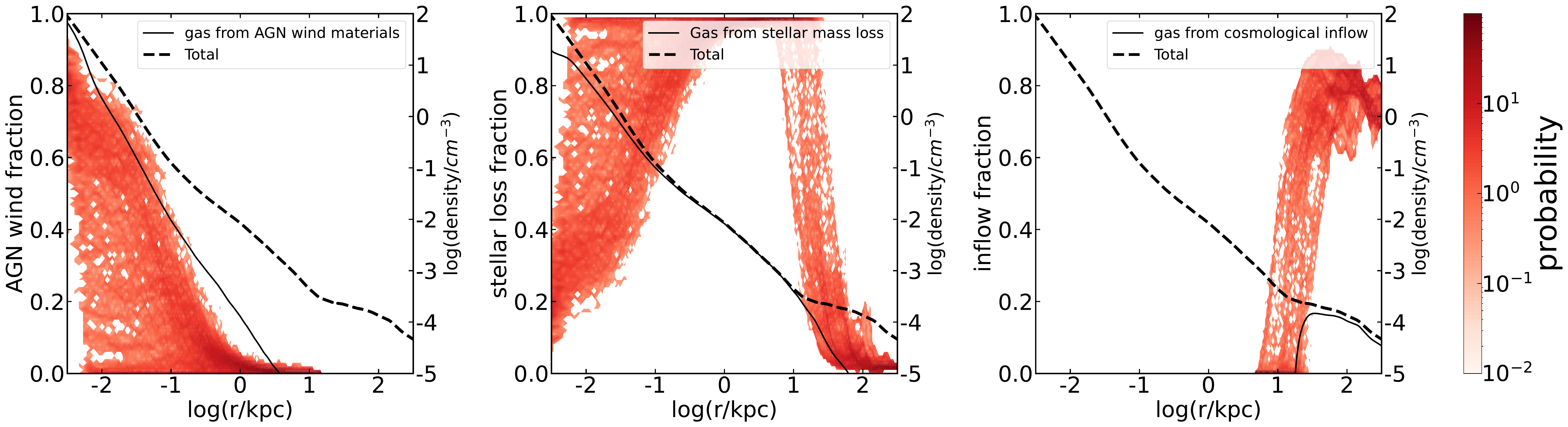}
    \caption{The radial distributions of the abundances of AGN wind, stellar mass loss, and cosmological inflow in the fiducial model stacked using simulation data at different times ranging from 7.5 to 12 Gyr. The depth of the color, therefore, represents the time probability distribution. The solid line in each plot represents the time- and $\theta$-averaged density of gas from AGN wind, stellar mass loss, and cosmological inflow, respectively. While the dashed line represents the total gas density. The simulation data from 7.5 to 12 Gyr is used for the time average.}
    \label{fraction}
   \end{figure*}

The fourth column of Figure \ref{snap_quiet} shows the distribution of the inflow abundance at 7.875 Gyr in the ``Fiducial'' model. The abundance is the gas ratio of the cosmological inflow to the total mass. The top panel shows that the cosmological inflow mixes with the ``original'' gas in the galaxy and forms a complex pattern. However, from the second panel, we can see that the cosmological inflow roughly stops at $\sim$20 kpc. In most regions outside 20 kpc, the cosmological inflow abundance is larger than 0.5, while within this radius, the abundance quickly drops to almost zero. 

To obtain an overall picture of the gas composition in the galaxy, we have drawn the radial distributions of the abundances of AGN wind, stellar loss abundance, and cosmological inflow using the simulation data from 7.5 to 12 Gyr. Figure \ref{fraction} shows the gas fraction, the average density originating from various sources and the total gas density in the ``Fiducial'' model. Here ``fraction'' is defined as the ratio of the gas mass from various sources, including AGN wind, mass loss from stellar evolution, and cosmological inflow, to the total gas mass at the given radius integrated overall $\theta$ values. For the initial gas, most of it has been expelled out of the halo, and the rest contributes around 20\% of the total gas outside 10 kpc. Since it is a subdominant component in the simulated galaxy, we ignore it in this work. Also shown in the figure is the time- and $\theta$-averaged number density of various gas components by the solid black lines and the number density of the total gas by the black dashed lines. It can be seen from the figure that the abundance of inflow is nearly as large as over $80\%$ at large radii, then drops sharply at $\sim$20 kpc, consistent with Fig. \ref{snap_quiet}. The stellar loss abundance is $\sim1$ at $1-$20 kpc, while within 1 kpc, the gas is the mixture of the AGN wind and the stellar loss. The abundance of AGN wind is over $90\%$ at $r\la 10^{-2}$ kpc, then gradually decreases with increasing radius. When the AGN wind is powerful, it can reach beyond 10 kpc. From Figure \ref{snap_quiet} and Figure \ref{fraction}, we can infer that the cosmological inflow can not enter the center of the galaxy and fuel the star formation and black hole accretion directly in massive galaxies. Almost all inflowing gas finally becomes part of circumgalactic media (CGM) and is blocked outside $\sim$20 kpc.
   
\begin{figure}
    \includegraphics[width=0.45\textwidth]{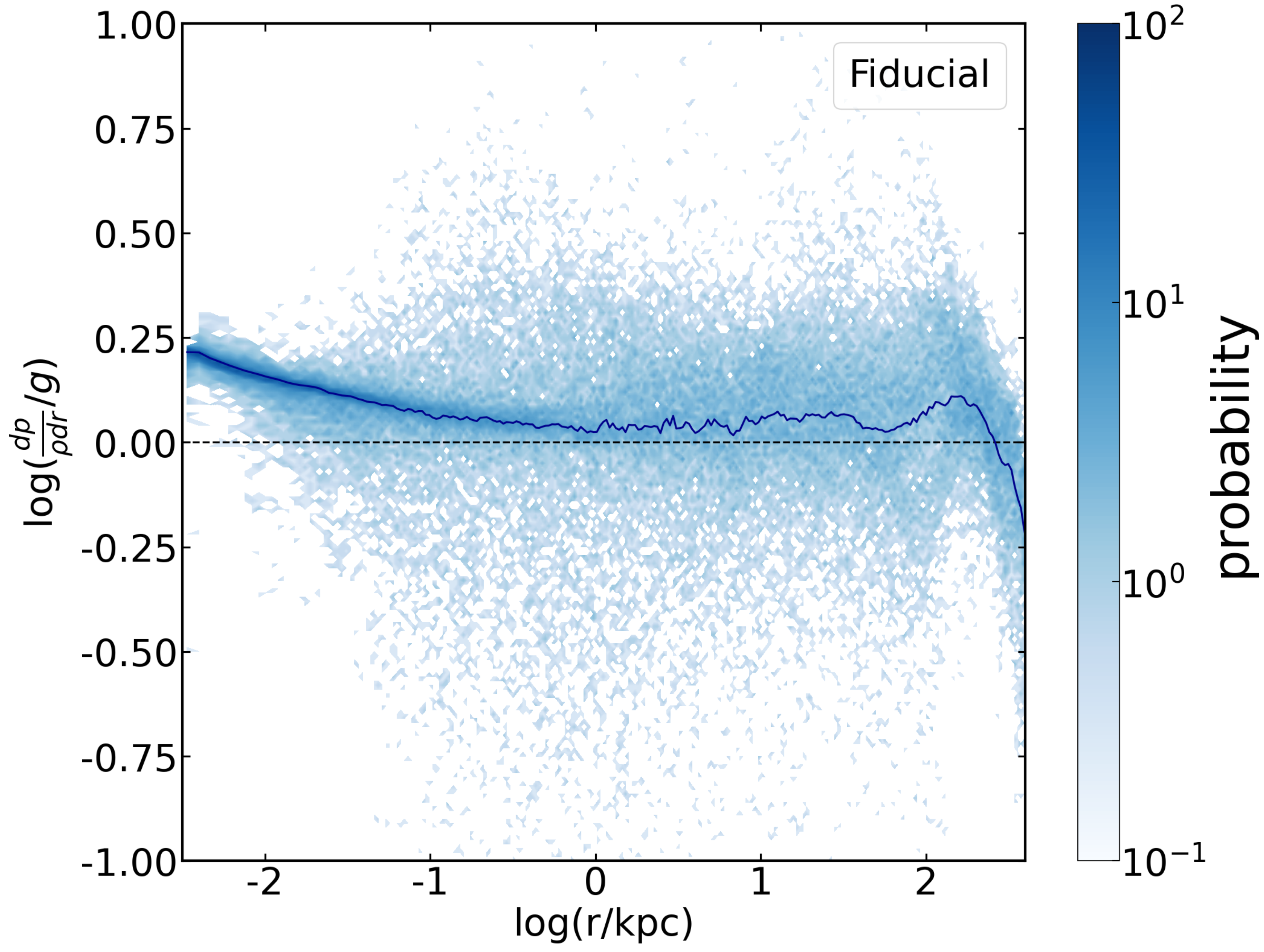}
    \caption{The stacked radial distribution of the ratio of radial gas pressure gradient and gravitational acceleration in the fiducial simulation using simulation data ranging from 7.5 to 12 Gyr. The depth of the color represents the time probability. The solid blue line represents the median value of the ratio, and the horizontal dashed black line represents the force ratio equal to 1.}
    \label{forceratio}
   \end{figure}

 \begin{figure}
    \includegraphics[width=0.45\textwidth]{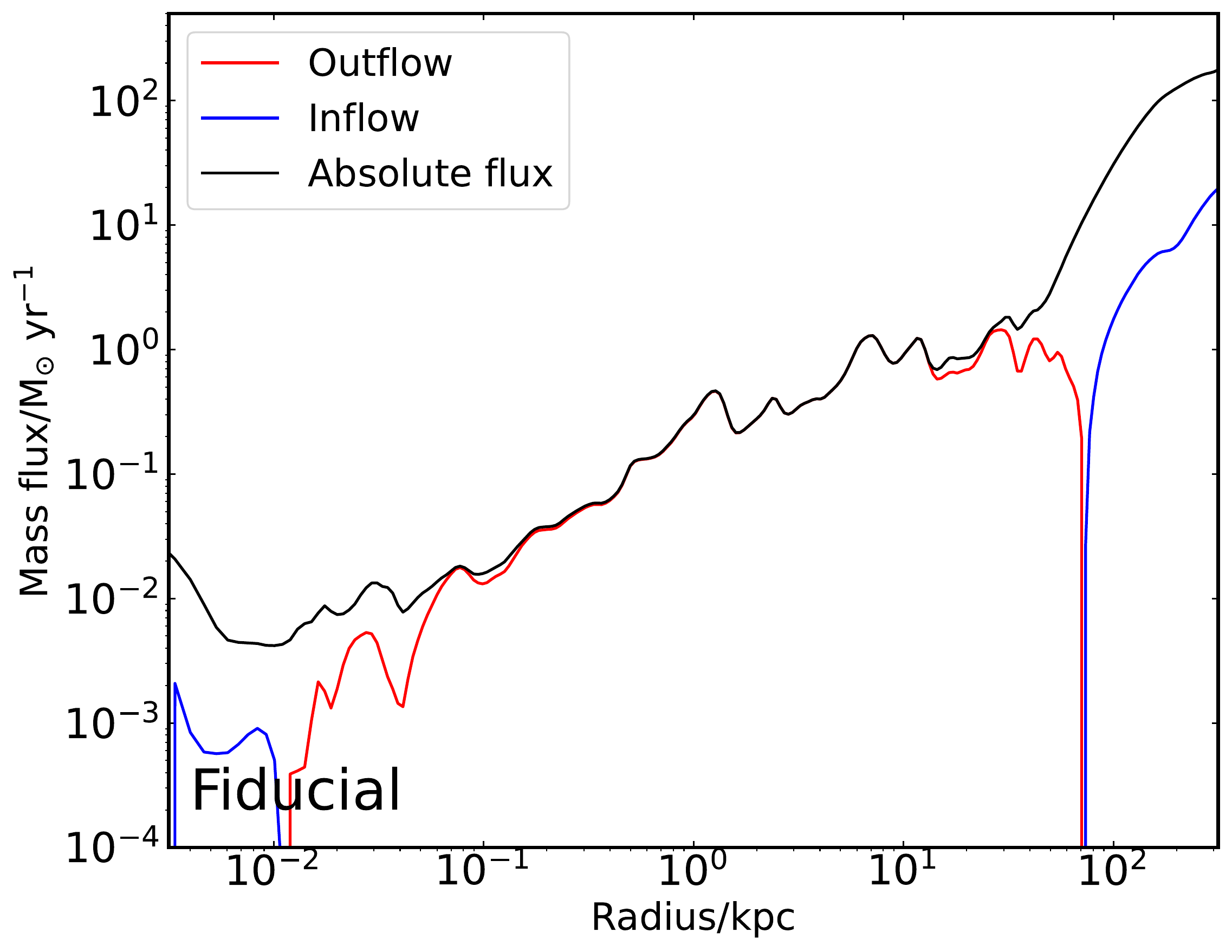}
    \caption{The time-averaged net radial mass flux (sum of radial flux through all cells at given radius) and absolute radial mixing flux (sum of the absolute value of radial flux through all cells) in ``Fiducial'' Model from 7.5 to 12 Gyr.}
    \label{radflux}
   \end{figure} 

\begin{figure*}
    \includegraphics[width=\textwidth]{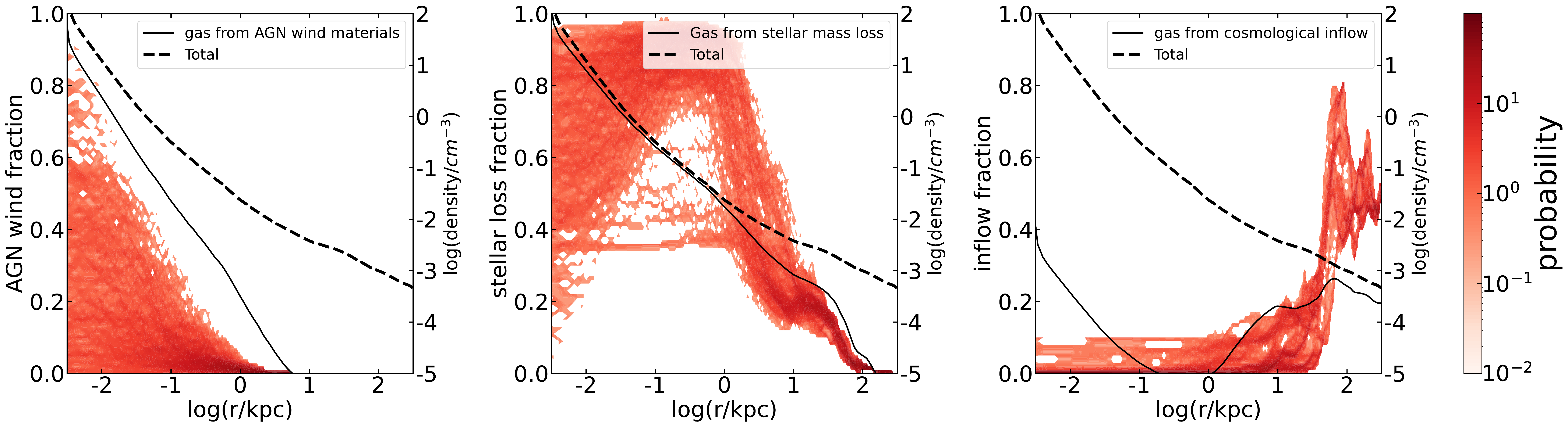}
    \caption{Similar to Figure \ref{fraction}, but for the ``3Inflow'' model.}
    \label{fraction_3inflow}
\end{figure*} 

\begin{figure*}
    \includegraphics[width=\textwidth]{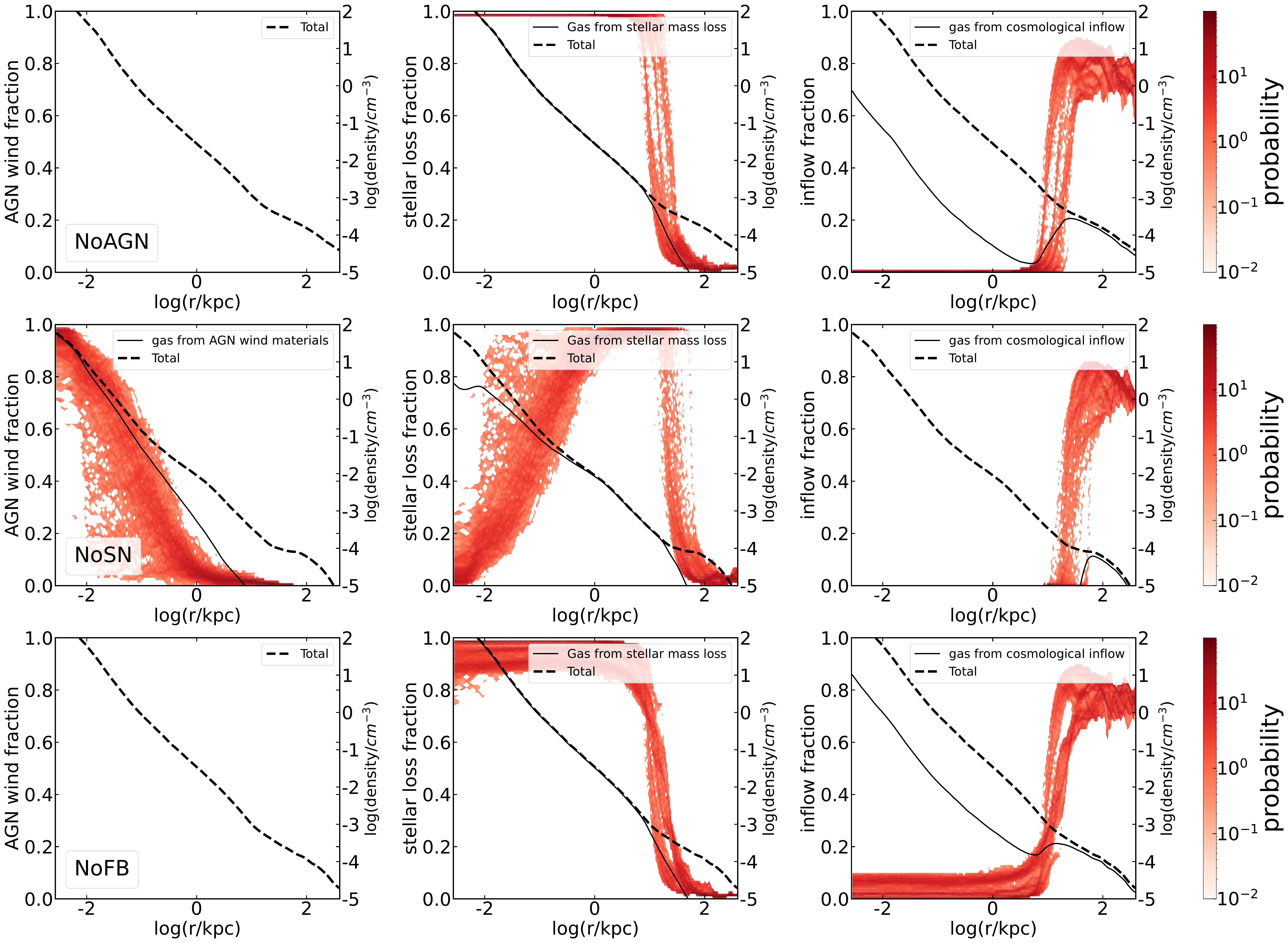}
    \caption{Similar to Figure \ref{fraction}, but for ``NoAGN'' (top), ``NoSN'' (middle), and ``NoFB'' (bottom) models.}
    \label{fraction_all}
\end{figure*} 

 \begin{figure*}
    \includegraphics[width=0.9\textwidth]{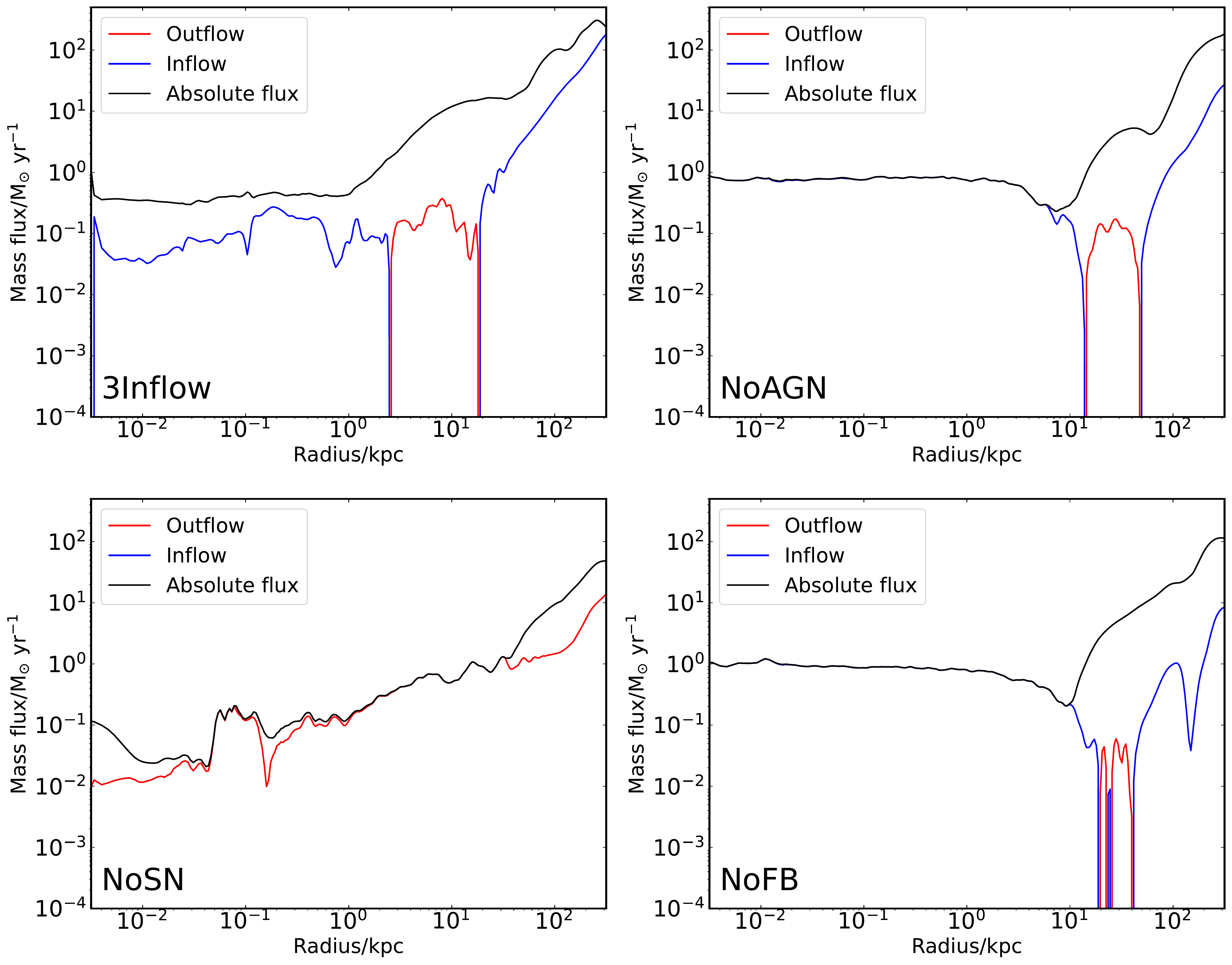}
    \caption{Similar to Figure \ref{radflux}, but for the ``3Inflow'', ``NoAGN'', ``NoSN'', and ``NoFB'' models.}
    \label{radflux_all}
   \end{figure*}

\subsection{Why can the cosmological inflow not enter the galaxy?}\label{why}

From the abundance of inflow gas shown in Figure \ref{snap_quiet}, we can see a clear boundary at $\sim20$ kpc between the inflow gas and the original gas in the galaxy. Why can the inflow not continue to fall within this radius? Physically, the reason why the inflow is stopped at a specific radius must be due to the force acting on the gas, combined with the ``initial velocity'' of the gas. Consider a cloud of cosmological inflow gas, and we neglect its angular momentum. Then two dominant forces acting on it are the gradient of gas pressure $\nabla P_{\rm gas}$, usually outward, and the gravitational force $\rho \nabla \Phi$. Here $P_{\rm gas}$ should come from the background gas surrounding the cloud, $\rho$ is the density of the cloud, and $\Phi$ denotes the gravitational potential. For simplicity, let us assume that the cloud's density equals the background gas. We have calculated the ratio of these two forces $\nabla P_{\rm gas}/\rho \nabla \Phi$ as a function of radius averaged over all $\theta$ angles. The results  at different times are shown in Figure \ref{forceratio}. The solid blue line represents the median value of the ratio. From the figure, we see that the ratio has a large scatter. For the median value, the ratio is smaller than one outside $\sim$200 kpc, i.e., the net force is inward at that region, implying that the inflow velocity keeps increasing in this region. Within $\sim$200 kpc, however, the gas pressure gradient becomes larger than the gravity, so the inflow decelerates in this region and should finally stop at a certain radius.

While the above force analysis can provide us a physical insight of why the inflow should stop at a certain region, it is difficult to obtain a quantitative estimation to the value of stop radius. For this aim, we have performed a convective stability analysis by calculating the entropy gradient. From the bottom-right panel of Figure \ref{dens&cooling}, We find that between $\sim$20 kpc and the outer boundary, the entropy gradient is flat, implying that the flow is convectively unstable and strong inflow-outflow motion should be present in this region\footnote{We note that mixing might be underestimated under our 2D axisymmetric simulation due to reduced surface mixing areas and needs to be better quantified in full 3D simulations in the future.}, which will determine the stop radius. This convection is likely driven by the AGN feedback, as have been pointed out by  \cite{yangreynolds2016} in the context of AGN feedback in a galaxy cluster.  The inflow thus can be carried by the convective motion and finally stops at $\sim$20 kpc. 

To clearly illustrate this scenario, Figure \ref{radflux} shows the time-averaged net radial mass flux, which is a sum of the radial flux through all cells at a given radius, as well as the absolute radial mixing flux, which is the sum of the absolute value of radial flux through all cells. The figure shows that the mass flux is negative outside $\sim$70 kpc and becomes positive inside $\sim$70 kpc, indicating that the ``overall'' inflow motion stops at this radius. This result is consistent with the analysis of Figure \ref{forceratio}. From the outer boundary until $\sim$20 kpc, the absolute flux is larger than the net mass flux, indicating radial mixing in this region due to convection.  This is why the stop radius is 20 kpc.

The force ratio depends on the density contrast between the inflow and the ``original'' gas in the galaxy, among other things. The analysis for Figure \ref{forceratio} assumes that the density of the inflow and the ``original'' gas is identical. The background gas is usually in hydrostatic equilibrium. Suppose the density of the inflow gas is higher than the background gas. In that case, we will have $\nabla P_{\rm gas}\la \rho \nabla \Phi$, so the inflow will keep moving inward until it stops at a radius where the density of the inflow is roughly equal to that of the surrounding background gas; so  we expect that the ``stop radius'' should be smaller if the density of the injecting inflow is higher. This is why cold, dense clumps can easily fall onto the center of the galaxy. 

To test this issue, we check the results of the ``3Inflow'' model. The higher mass flux in this model is achieved by increasing the mass density of the inflow at the outer boundary of our simulation domain. Similar to Figure \ref{fraction}, in Figure \ref{fraction_3inflow}, we draw the radial distribution of the abundances of AGN wind, stellar loss, and cosmological inflow within 7.5-12 Gyr for the ``3Inflow'' model, together with the density profiles of each component and their sum. Compared with the results of the ``Fiducial'' model, the fraction of inflowing gas in $1-10$ kpc can reach $\sim10\%$, significantly higher than the ``Fiducial'' model. 

Similar to Figure \ref{radflux}, we have also calculated the radial fluxes for the ``3Inflow'' model.  The top-left panel of Figure \ref{radflux_all} shows the time-averaged net radial mass flux and absolute radial mass flux for the ``3Inflow'' model. Similar to Figure \ref{radflux}, we can also see that the net mass flux has a similar structure between the ``3Inflow'' model and the ``Fiducial'' model. However, the inflow region at the outer part of the halo in the ``3Inflow'' model is larger than in the ``Fiducial'' model, with the boundary of inflow and outflow being 20 kpc in the ``3Inflow'' model. This result is consistent with our analysis that the higher inflow rate with higher density will make the cosmological inflow goes deeper in the galaxy. We also note that the absolute mass flux is larger than the net mass flux at all radii, implying that the radial mixing will take the cosmological inflow to reach the center of the galaxy. This is consistent with the result of Figure \ref{fraction_3inflow}.

\subsection{The roles of AGN and stellar feedbacks on blocking the cosmological inflow}\label{infb}

In section \ref{why}, we show that the gradient force of gas pressure is the main force that balances the gravitational force and prevents the cosmological inflow from entering the galaxy. The gas pressure depends on the distributions of density and temperature of the gas, while the temperature is determined by the heating and cooling processes, among other things. SN Ia and AGN feedback are two dominant heating mechanisms. In this section, we investigate whether they are essential in constructing the gas pressure and blocking the cosmological inflow.

For this aim, similar to Figure \ref{fraction}, we first calculate the gas fraction of various gas sources in the ``NoAGN'', ``NoSN'', and ``NoFB'' simulations and compare their results with the ``Fiducial'' model. We then can know whether the cosmological inflow can enter deeper into the galaxy when the AGN and SN feedback processes are absent. Figure \ref{fraction_all} shows the results. Similar to the ``Fiducial'' model, we can see from the figure that the cosmological inflow is also blocked at certain radii. The stop radius in the ``NoAGN'' and ``NoFB'' models is $\sim$ 10 kpc while it is  $\sim30$ kpc in the ``NoSN'' model. 

In the ``NoFB'' model the fraction of the cosmological inflow can reach $\sim10\%$ at the inner region from $r\sim10$ kpc to the inner boundary of the simulation domain. In this model, $\sim90\%$ of the gas comes from the stellar mass loss. In the ``NoAGN'' model, this fraction decreases significantly, while in the ``NoSN'' model this fraction almost decreases to zero. 

Comparing the gas density in the two models without AGN (i.e., ``NoAGN'' and ``NoFB'' models) shown in Figure \ref{fraction_all} and the ``Fiducial'' model shown in Figure \ref{fraction}, we can see that the density in the former is several times higher than in the latter at the inner region of the galaxy. This is, of course, because of the absence of heating by AGN in the ``NoAGN'' and ``NoFB'' models. On the other hand, we note that, in these two models, even though there is no heating from AGN, the gas in the galaxy remains roughly hot, and no catastrophic cooling occurs. To understand this result, we calculated the gas cooling timescale at different radii and compared it with the local inflow timescale. We find that the former is always larger, implying that the gas does not have enough time to cool down. Physically, the long cooling timescale is because the gas density is still low. The inflow timescale is relatively short partly due to the small angular momentum of the gas in the simulated galaxy.  

To understand the values of the stop radius in different models shown in Figure \ref{fraction_all}, following Figure \ref{radflux}, we have  calculated the time-averaged net radial and absolute mass fluxes in the ``NoAGN'', ``NoSN'', and ``NoFB'' models. The results are shown in Figure \ref{radflux_all}. We can see from the figure that the ``NoAGN'' and ``NoFB'' models are similar. Outside $\sim$50 kpc, it is an overall inflow region (i.e., the net flux is negative), followed by an overall outflow region (i.e., the net flux is positive) with decreasing radius and then an inflow region again.  Outside $\sim$10 kpc, the absolute flux is larger than the net radial flux in these two models, implying strong convection in this region. This explains why the values of the stop radius of cosmological inflow in these two models are both $\sim$10 kpc. 
Different from these two models, the net  mass flux in the ``NoSN'' model is positive at all radii. This is because the AGN in this model is much stronger, triggering stronger outflow. 
Outside 30 kpc the absolute mass flux  is larger than the net radial mass flux, implying convection there and explaining the stop radius of 30 kpc shown in Figure 
Figure \ref{fraction_all}. At last, it is interesting to note that a convection region within $\sim100$ pc that is present in both the ``Fiducial'' and ``NoSN'' models disappears in the ``NoAGN'' and ``NoFB'' models, implying that the convection is triggered by AGN feedback.

The above results suggest that the gas pressure to block the cosmological inflow mainly comes from the thermalized stellar wind instead of SN and AGN feedback. First, the temperature of the thermalized stellar wind is determined by the Jeans equation, the typical temperature of thermalized stellar wind is around the virial temperature, and the thermal energy of the thermalized stellar wind is comparable to its gravitational energy. More importantly, the stellar wind is the dominant gas source in the galaxy. The continuous materials supply provides a stable hot gas core at the center, blocking the cosmological inflow.

It is worth considering which has a more significant impact on preventing inflow - SN or AGN feedback - even though both have a smaller effect than stellar wind. Figure \ref{fraction_all} shows that the inflow fraction in the ``NoSN'' model is very similar to that in the ``Fiducial'' model. However, the average density of the gas from cosmological inflow in the inner region in the ``NoAGN'' model is higher than that in the ``NoSN'' and ``Fiducial'' models. This result implies that the AGN feedback is the secondarily important mechanism to block the cosmological inflow in addition to the thermalised stellar wind. To understand this result, we have calculated the total energy emitted by the AGN and the SN for the ``Fiducial'' model during the whole evolution period. We find that the former is roughly one order of magnitude larger than the latter\footnote{This result will be further discussed in Zhu et al. (2023, in preparation).}. From the second row of Figure \ref{fraction_all}, we can also see that AGN wind can propagate to $\sim80$ kpc. Considering the shock produced by the wind, the energy produced by the AGN can be transported even further. While for the stellar feedback, its energy can only deposit within $\sim r_{\rm e}$, which is $\sim7$ kpc in our simulations. Given the higher energy output and farther energy transport distance, we can understand why the AGN feedback significantly impacts heating the gas compared to SN feedback. 

\subsection{The effects of cosmological inflow on SFR and BHAR}\label{effects}

\begin{figure*}
    \includegraphics[width=0.95\textwidth]{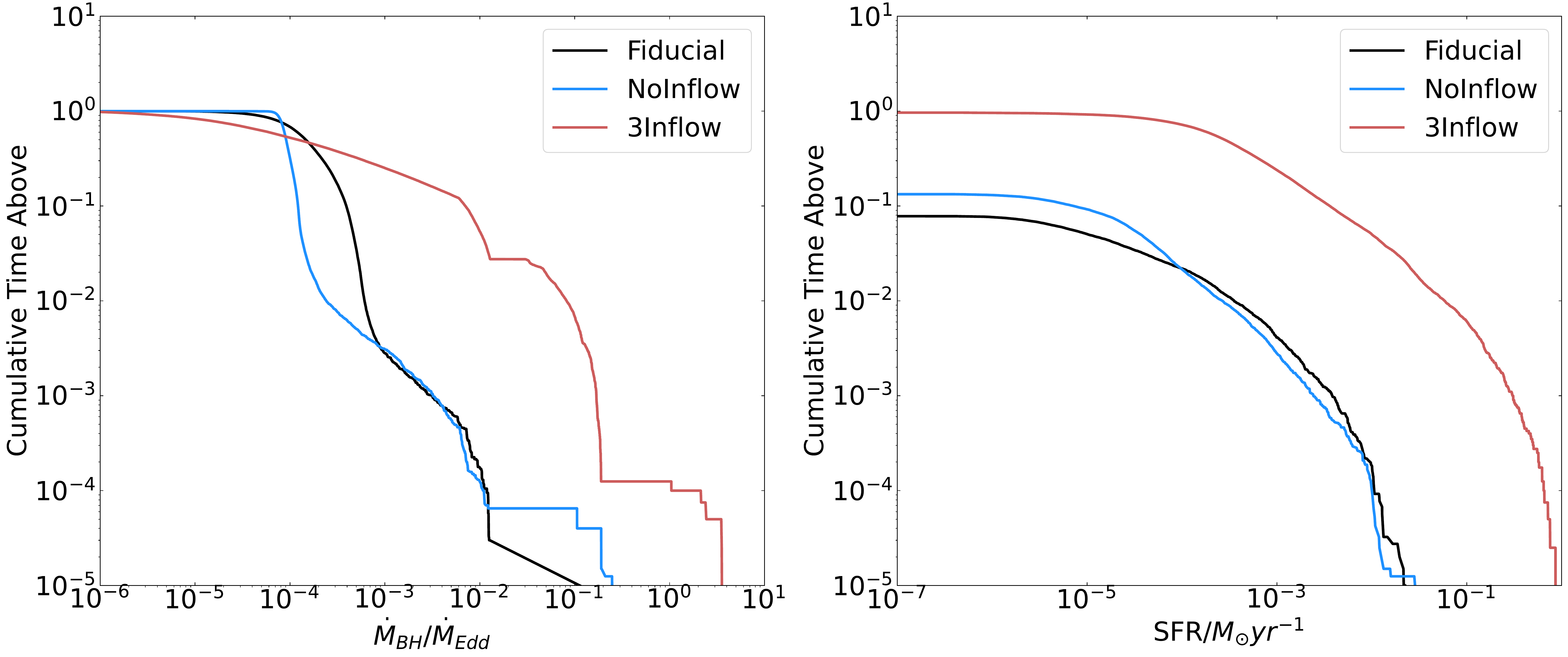}
    \caption{The fraction of the cumulative time above the corresponding star formation rate and black hole accretion rate in the ``Fiducial'', ``NoInflow'', and ``3Inflow'' models from 7.5 to 12 Gyr.}
    \label{dutycycle}
\end{figure*}

\begin{figure*}
    \includegraphics[width=0.95\textwidth]{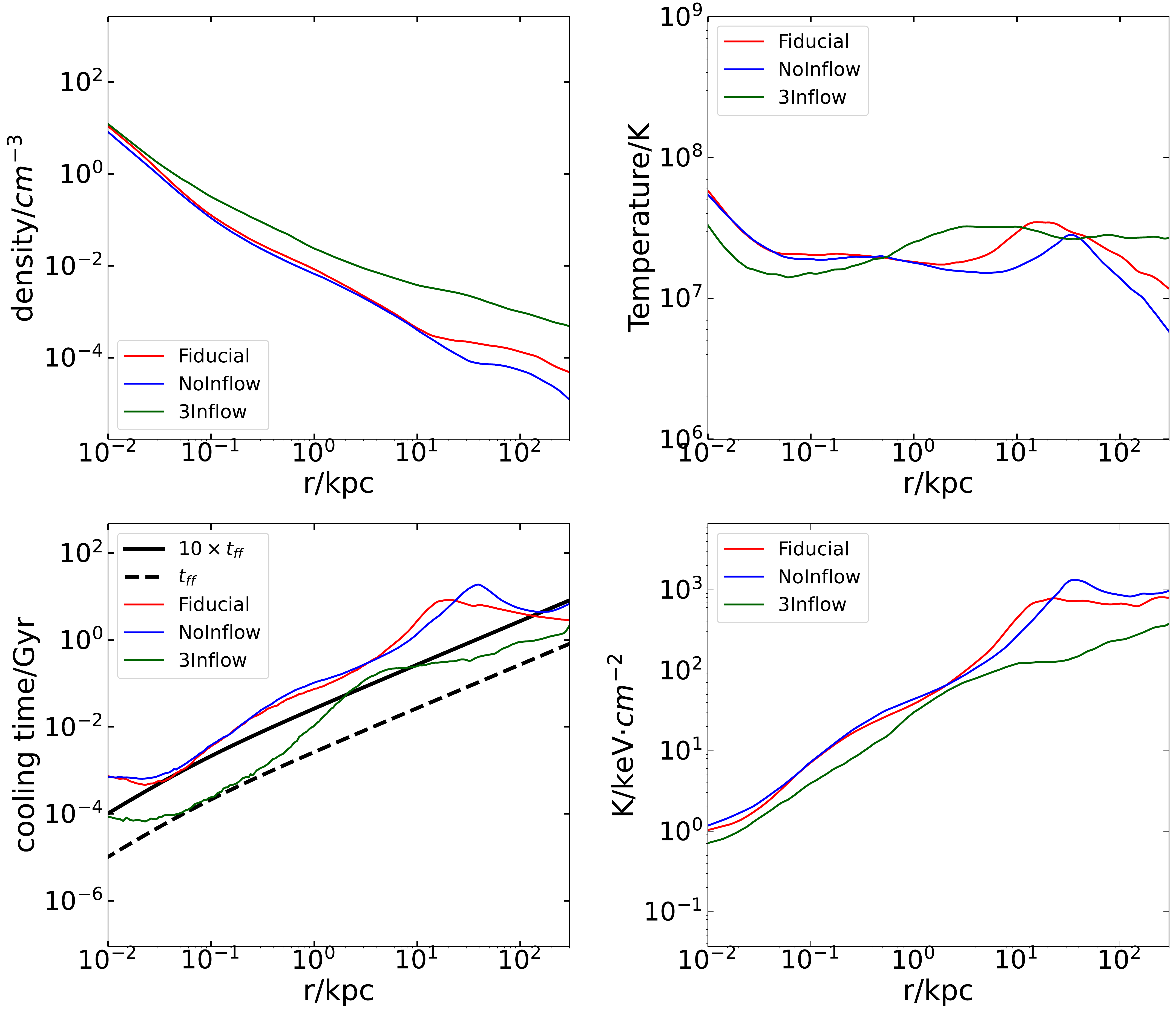}
    
    \caption{The median radial distribution of number density, temperature, cooling time, and entropy in the ``Fiducial'', ``NoInflow'', and ``3Inflow'' models. The values are obtained by calculating the median value using  the simulation data from 7.5 to 12 Gyr.}
    \label{dens&cooling}
\end{figure*}

It is often assumed that cosmological inflow is a vital gas source for star formation and black hole accretion \citep{SA2014, DeGraf2017}. We now quantitatively evaluate the effects of cosmological inflow on SFR and BHAR. We do this by comparing the values of SFR and BHAR in the ``Fiducial'', ``NoInflow'', and ``3Inflow'' models, shown in Figure \ref{SFR&BHAR}. We find that the BHAR and SFR are roughly the same for the ``Fiducial'' and ``NoInflow'' models. The main reason is that the cosmological inflow hardly enters the region within $\sim$20 kpc in both simulations, while star formation mainly occurs in the central region of the galaxy. However, in the ``3Inflow'' model, we find that the inflow significantly enhances both the SFR and BHAR compared to the ``Fiducial'' model, with the BHAR and SFR increased by $\sim$5 times and 3 orders of magnitude, respectively. 

To obtain more quantitative results, we further show the fraction of cumulative time above the BHAR and SFR of these three models from 8 Gyr to 12 Gyr in Figure \ref{dutycycle}. The figure shows that the duty cycle of BHAR and SFR in the ``Fiducial'' and ``NoInflow'' models are similar. Less than $0.01\%$ of the time, the values of BHAR in both models are above $10^{-2}~\dot{M}_{\mathrm{Edd}}$, and the AGN enters the cold mode. Only around $10\%$ of the time, the values of SFR in both models are greater than $10^{-7}~\mathrm{M_{\odot}\ yr^{-1}}$. However, the duty cycle of the ``3Inflow'' model is significantly different from the ``Fiducial'' and ``NoInflow'' models. In the ``3Inflow'' model, the BHAR is larger than $10^{-2}~\dot{M}_{\mathrm{Edd}}$ at $>1\%$ of the time, while almost all of the time, the SFR is larger than $10^{-7}~\mathrm{M_{\odot}\ yr^{-1}}$. We conclude that the cosmological inflow can significantly enhance SFR and BHAR if the inflow rate is large enough. 

To investigate why the BHAR and SFR are high in the ``3Inflow'' model, we have shown the median radial profile of gas number density, temperature, cooling time, and entropy for ``NoInflow'', ``Fiducial'' and ``3Inflow'' models using their simulation data from 7.5 to 12 Gyr from in Figure \ref{dens&cooling}. The figure shows that the profiles of these four quantities within $\sim$10 kpc are very similar for the ``Fiducial'' and ``NoInflow'' models. Based on \cite{sharma12}, the thermal instability will be triggered when the cooling timescale is shorter than 10 times the free-fall timescale in the hot halo. However, in both models, we can see from the figure that cooling time does not fall below 10 times free-fall timescale within 10 kpc. Although out of 10 kpc, the gas density in the ``Fiducial'' model is higher than the ``NoInflow'' model due to cosmological inflow in the former, the thermal state of the gas in the inner region does affect the star formation and black hole accretion. Since the gas density and cooling time of the ``Fiducial'' and ``NoInflow'' models within 10 kpc are similar, the SFR and BHAR are thus similar in these two models.

On the other hand, the gas density in the ``3Inflow'' model is about 20-30 times higher than the other two models. Correspondingly, in this model, the cooling timescale is shorter than 10 times the free-fall timescale, so we expect the significant formation of cold gas in this model. This result explains why the SFR and BHAR are significantly higher than the other two models. 

What is the reason for enhancing density in the galaxy in the ``3Inflow'' model? We would like to point out that density enhancement is not because of the additional material supply from the cosmological inflow. From Figure \ref{fraction_3inflow}, we can see that the inflowing gas fraction within $\sim$10 kpc is only $\sim$20\%, so the gas in the galaxy is dominated by the stellar mass loss, as we have emphasized before. This question has been investigated in \cite{voit20}. \cite{voit20} have proposed that the CGM pressure at the outer boundary will enhance the central pressure and density in the galaxy. Mathematically, this is because the outer boundary conditions will determine the solution of differential equations to some degree. The gas density (and pressure) at the outer region in the ``3Inflow'' model is 3 times higher than the other two simulations, which will increase the gas density in the galaxy. Moreover, such an increase in gas density will result in stronger cooling, further increasing the gas density. This explains why the gas density in the whole galaxy in this model is 20-30 times higher than in the other two models. Then why are the gas density in the ``Fiducial'' and ``NoInflow'' models similar, although the density in the ``Fiducial'' model is higher than the ``NoInflow'' model at the outer boundary? \cite{voit20} argue that this is because a critical value of density (or pressure) exists at the outer boundary. Only when the density at the outer boundary is higher than this value does the outer boundary condition significantly affect the density in the galaxy. The existence of this critical value is because thermalized stellar wind provides the main source of gas pressure in the galaxy. A high gas density will make the gas vulnerable to thermal instability and the formation of cold gas \citep{sharma12} and stars. This is why the SFR and BHAR in the ``3Inflow'' model are the highest.

\begin{figure}

\includegraphics[width=0.45\textwidth]{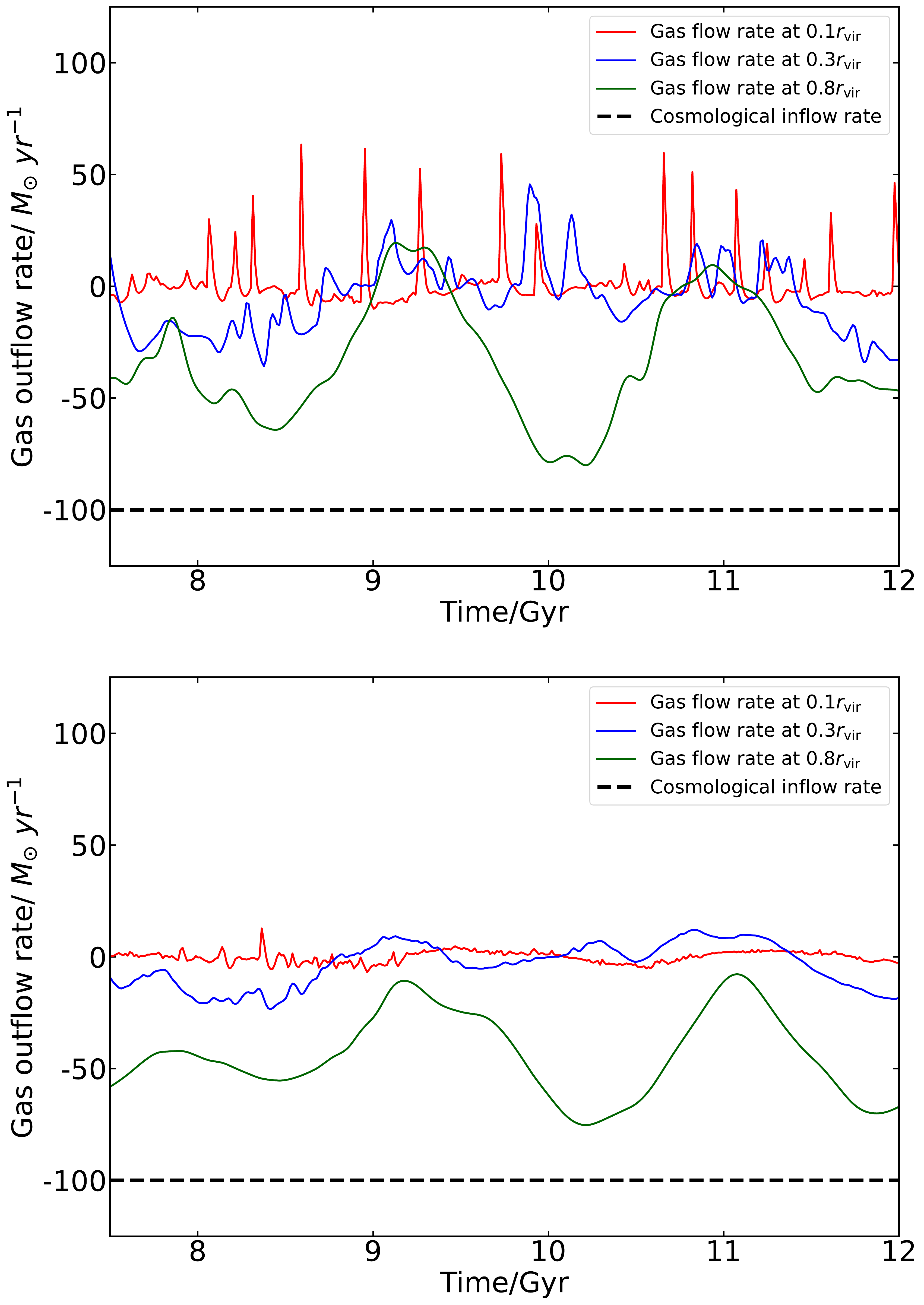}\\
    \caption{The time evolution of net flow rate of the gas coming from cosmological inflow at different radii from 6 to 12 Gyr in the ``Fiducial'' ({\it top panel}) and ``NoAGN'' ({\it bottom panel}) models. The value of the gas flow rate is averaged with 125 Myr. The red, blue, and dark green lines represent the net flow rate at 0.1$r_{\mathrm{vir}}$, 0.3$r_{\mathrm{vir}}$, and 0.8$r_{\mathrm{vir}}$, respectively. Positive values represent outflow, while negative ones represent inflow. The dashed line represents the cosmological inflow rate we set at the outer boundary in these two models. }
    \label{netrate}
\end{figure}

\begin{figure}
    \includegraphics[width=0.45\textwidth]{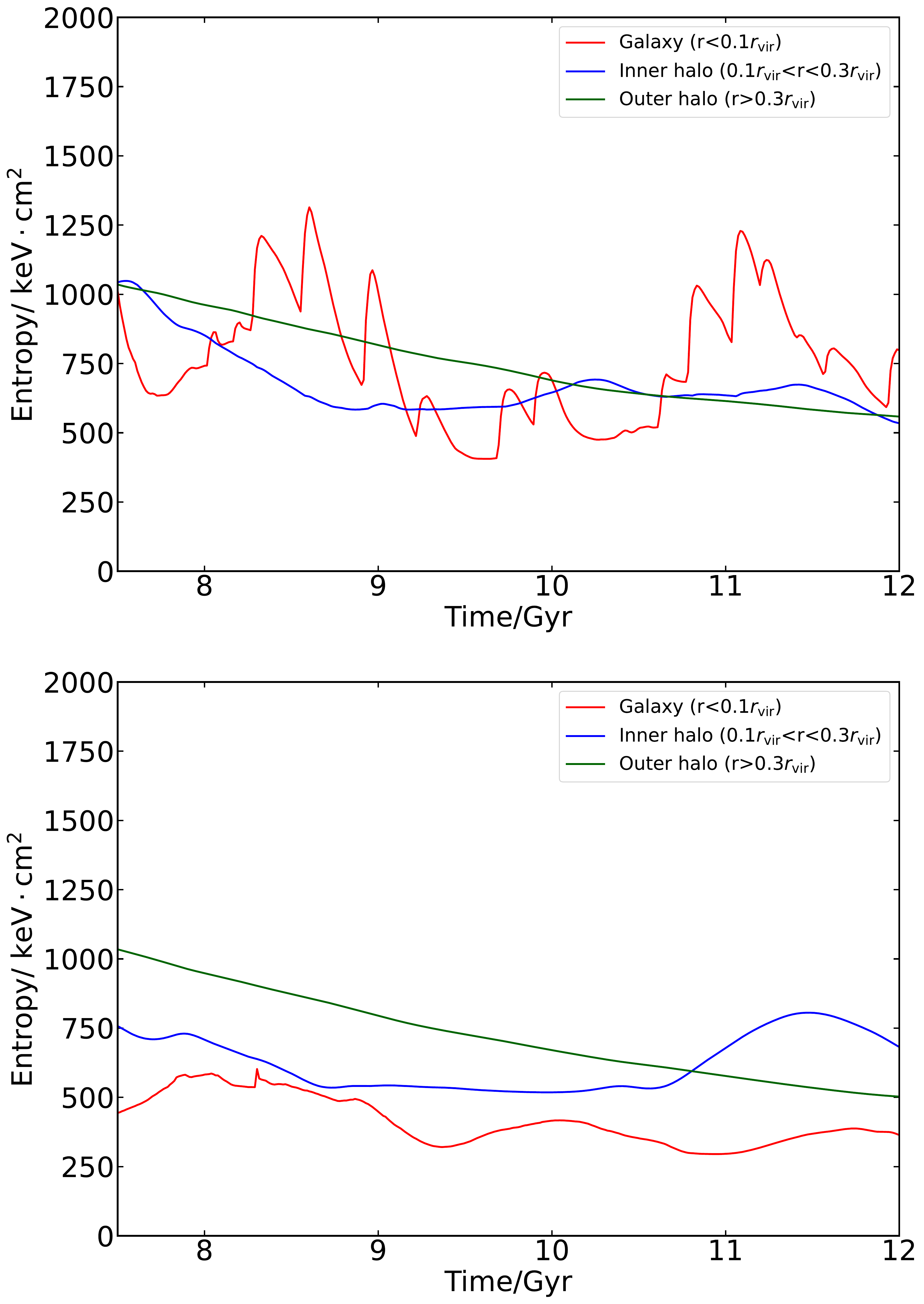}\\

    \caption{Time evolution of gas entropy in different regions in the ``Fiducial'' ({\it top panel}) and ``NoAGN'' ({\it bottom panel}) models. The red, blue, and dark green lines represent the gas entropy within 0.1$r_{\mathrm{vir}}$, between 0.1$r_{\mathrm{vir}}$ and 0.3 $r_{\mathrm{vir}}$, and outside 0.3$r_{\mathrm{vir}}$, respectively. }
    \label{timeentropy}
\end{figure}

\subsection{The influences of AGN feedback on  CGM}\label{AGN&inflow}

Some previous works have studied the effects of  AGN feedback on the cosmological inflow \citep{nelson15, somerville15,correa18b, zinger20}. For example, \citet{zinger20} has analyzed the TNG data and found that AGN feedback can heat the inflow. In this section, in {\it MACER} framework, we also investigate the influence of AGN feedback on cosmological inflow, which is accumulated in the CGM region. 

For this aim, we have first compared the evolution of the net rate of the gas coming from the cosmological inflow at three different radii in the ``Fiducial'' and ``NoAGN'' models. Figure \ref{netrate} shows the results. We can see from the figure that, in both models, most of the time, the inflow is really inflowing (since it is negative). The inflow rate decreases inward. Taking the ``Fiducial'' model as an example, the rates at 0.1~$r_{\rm vir}$, 0.3~$r_{\rm vir}$, and 0.8~$r_{\rm vir}$ averaged in 7.5-12 Gyr are -0.32~${\rm M_{\odot}/yr}$, -6.26~${\rm M_{\odot}/yr}$, and ${\rm -43~M_{\odot}/yr}$, respectively\footnote{If we also take into account all component of the gas, i.e., AGN wind, cosmological inflow, stellar mass loss, and initial gas, the net flow rates at 0.1~$r_{\rm vir}$, 0.3~$r_{\rm vir}$, and 0.8~$r_{\rm vir}$ will be 1.08~${\rm M_{\odot}/yr}$, -4.99~${\rm M_{\odot}/yr}$, and ${\rm -35~M_{\odot}/yr}$.}. The rapid decrease of the inflow rate with decreasing radius result again indicates that only a small fraction of the cosmological inflow can reach a small radius, and most of them are accumulated in the CGM region, consistent with Figure \ref{fraction}. In the case of the ``Fiducial'' model, we can find the intermittent occurrence of the outburst of outflow. This is most obvious at 0.1~$r_{\rm vir}$. Such outbursts are almost absent in the case of the ``NoAGN'' model. The duration of each outburst increases with increasing radii, suggesting that the outbursts are driven at small radii and then propagate outward. Detailed cross-correlation analysis of the three light curves should be able to reveal the time lag between the outbursts at three radii, which is beyond our present work and could be analyzed in future work. Since the cosmological inflow rate we set is constant, such variation in gas flow rate and the galactic outflows must be triggered by strong AGN activities. The presence of some peaks in the blue line of the ``Fiducial'' model seems to indicate that AGN feedback can affect the cosmological inflow as far as $\ga 0.3~r_{\rm vir}$ by its intermittent outbursts.

To study whether the AGN feedback can transfer energy to the CGM region, we have calculated the time evolution of gas entropy density in three different regions of the galaxy in the ``Fiducial'' and ``NoAGN'' models, namely the ``galaxy'' (red line), ``inner halo'' (blue line), and ``outer halo'' (black line) regions. The results are shown in Figure \ref{timeentropy}. We can see from the figure that, within 0.1~$r_{\mathrm{vir}}$ (i.e., in the ``galaxy'' region), the entropy in the ``Fiducial'' model is higher and has much stronger oscillations compared with the ``NoAGN'' model. Within $0.1~r_{\mathrm{vir}}-0.3~r_{\mathrm{vir}}$ (i.e., the ``inner halo'' region), the entropy in the ``Fiducial'' model is also significantly larger than the ``NoAGN'' model most of the time\footnote{By checking the simulation data, we find that the increase of entropy in the ``NoAGN'' model in inner halo at the final 1 Gyr is due to a high-entropy clump flowing into this region.}. Outside 0.3~$r_{\mathrm{vir}}$ (i.e., the ``outer halo'' region), the time evolution of entropy in both models is almost the same. These results indicate that the AGN feedback can heat the gas outside 0.1~$r_{\mathrm{vir}}$. 

The above results are consistent with our previous analysis of AGN wind abundance shown in Figure \ref{fraction}. The left panel of Figure \ref{fraction} shows that the AGN wind can reach $\sim$80 kpc, which is beyond  0.1~$r_{\mathrm{vir}}\sim50$ kpc, thus can transport the AGN energy at least to that radius. Since the AGN wind cannot reach 0.3 $r_{\mathrm{vir}}$, this explains why the entropy of the outer halo of the ``Fiducial'' and ``NoAGN'' models shown in Figure \ref{timeentropy} are almost the same. 

\section{Summary}\label{summary}

In this paper, we have performed a two-dimensional high-resolution hydro-dynamical simulation to investigate the fate and impacts of cosmological inflow in the elliptical galaxy in the framework of {\it MACER}. In this framework, the inner boundary of the simulation is small enough to resolve the Bondi radius so that the accretion rate of the central black hole can be reliably determined, and state-of-the-art AGN physics is adopted. The time evolution data of the physical properties of the cosmological inflow is extracted from the IllustrisTNG cosmological simulations (Figure \ref{tng_inflow}) and implemented in our simulation at the outer boundary.

We start our simulation at 2 Gyr after the big bang when the target elliptical galaxy is already formed. To understand the interplay between AGN and SN feedback and cosmological inflow, in addition to the ``Fiducial'' model in which both AGN and SN feedback are included, and the cosmological inflow has been considered, we have also run five additional models for comparison purposes, including  ``NoAGN'' (AGN is not included), ``NoSN'' (SN is not included), ``NoFB'' (both AGN and SN are not included), ``NoInflow'' (cosmological inflow is not included) and ``3Inflow'' (3 times inflow rate is adopted) models. Our main results can be summarized as follows.
\begin{enumerate}[1.]

\item In the ``Fiducial'' model, the cosmological inflow falls into the galaxy from the outer boundary but is then blocked at $\sim$20 kpc and becomes part of the CGM (Figures  \ref{snap_quiet} \& \ref{fraction}). For the ``3Inflow'' model, some fraction of the cosmological inflow can enter the galaxy, although most cannot. In the region of $\la 10$ kpc, about 10\% of the gas at each radius comes from the inflow (Figure \ref{fraction_3inflow}). 

\item The physical reason for stopping the inflow from falling further is found to be  the gradient force of gas pressure in the galaxy (Figure \ref{forceratio}). An quantitative estimation to the stop radius of the inflow can be obtained by considering the radial mixing due to convective motion which is present outside $\sim$20kpc (Figure \ref{radflux}).

\item  Neither AGN feedback nor SN feedback is found to be the dominant process for preventing the cosmological inflow from entering the galaxy. Instead, the thermalized stellar mass loss provides the gas pressure. The stellar wind provides the main source of gas in the galaxy while the temperature of the thermalized stellar wind has the local virial value.

\item By comparing the inflow fraction in the ``NoAGN'' and ``NoSN''models (Figure \ref{fraction_all}), we find that AGN feedback is the secondarily important mechanism for blocking the cosmological inflow after the stellar mass loss, i.e., it is more important than SN feedback. Physically, this is because the total energy released from AGN is about one order of magnitude larger than SN feedback, and the maximum distance the AGN feedback energy can be transported is $\ga 80$ kpc, which is much larger than that of  SN feedback. 

\item  Although AGN feedback is not the dominant process of blocking the cosmological inflow, its intermittent outburst can affect the gas in the galaxy as far as $\ga 0.3 r_{\rm vir}$, in terms of affecting the motion of the gas (Figure \ref{netrate}) and transport energy of the AGN outburst to increase the entropy of the gas (Figure \ref{timeentropy}). 

\item By comparing the ``Fiducial'' and ``NoInflow'' models, it is found that both the star formation rate and the black hole accretion rate are not strongly affected by the inclusion of the cosmological inflow (Figure \ref{SFR&BHAR}).

\item However, compared to the ``Fiducial'' model, the BHAR and SFR in the ``3Inflow'' model are increased by $\sim$5 times and 3 orders of magnitude respectively (Figure \ref{SFR&BHAR}). This is because the inflow in the ``3Inflow'' model makes the gas density at the CGM region higher than a critical value, which is determined by the gas supply by the stellar evolution so that the gas density in the galaxy is increased by a factor of  20-30. The gas density in the CGM region in the ``Fiducial'' model is lower than this critical value, so the gas density in the galaxy is not affected. Higher gas density makes the hot gas more vulnerable to thermal instability and enhances the formation of cold gas and the values of SFR and BHAR.  

\end{enumerate}

\section*{Acknowledgments}
We thank Prof. Volker Springel for the useful discussions, and the anonymous referee for his/her constructive suggestions and comments, which have significantly improved the paper. BZ, FY, and SJ are supported in part by the NSF of China (grants 12133008, 12192220, and 12192223), YP by the NSF of China (grants  12125301, 12192220, and 12192222) and the science research grants from the China Manned Space Project (CMS-CSST-2021-A07), LCH by the NSF of China (grants 11721303, 11991052, 12011540375, and 12233001) and the China Manned Space Project (CMS-CSST-2021-A04, CMS-CSST-2021-A06).  The calculations have made use of the High Performance Computing Resource in the Core Facility for Advanced Research Computing at Shanghai Astronomical Observatory. 
\section*{Data Availability}
The data underlying this article will be shared on reasonable request to the corresponding author.

\bibliography{ref}

\end{document}